\documentclass[fleqn,usenatbib]{mnras}

\usepackage{newtxtext,newtxmath}

\usepackage[T1]{fontenc}

\DeclareRobustCommand{\VAN}[3]{#2}
\let\VANthebibliography\thebibliography
\def\thebibliography{\DeclareRobustCommand{\VAN}[3]{##3}\VANthebibliography}
\newcommand{\ww}[1]{{\textcolor{black}{#1}}}

\usepackage{graphicx}	
\usepackage{amsmath}	
\usepackage{subfigure}
\usepackage{threeparttable}

\title[Thermal emission of WASP-103 b]{THERMAL EMISSION FROM THE HOT JUPITER WASP-103\,b IN $J$ AND $K{\rm s}$ BANDS}

\author[Y. Shi et al.]{
Yaqing Shi,$^{1,2,3}$
Wei Wang,$^{1,3}$\thanks{E-mail: wangw@nao.cas.cn}
Gang Zhao,$^{1}$
Meng Zhai,$^{1,3}$
Guo Chen,$^{4}$
Zewen Jiang,$^{1,2}$
\newauthor
Qinglin Ouyang,$^{2,3}$
Thomas Henning,$^{5}$
Jingkun Zhao,$^{1}$
Nicolas Crouzet$^{6}$
\newauthor
and Roy van Boekel,$^{5}$
\\
$^{1}$CAS Key Laboratory of Optical Astronomy, National Astronomical Observatories, Chinese Academy of Sciences, Beijing 100101, PR China;\\
$^{2}$School of Astronomy and Space Science, University of Chinese Academy of Sciences, Beijing 101408, China;\\
$^{3}$Chinese Academy of Sciences South America Center for Astronomy, Chinese Academy of
Sciences, Beijing 100012, PR China;\\
$^{4}$Key Laboratory of Planetary Sciences, Purple Mountain Observatory, Chinese Academy of Sciences, Nanjing 210023, PR China;\\
$^{5}$Max Planck Institute for Astronomy, K$\ddot{o}$nigstuhl 17, D-69117 Heidelberg, Germany;\\
$^{6}$Leiden Observatory, Leiden University, Postbus 9513, 2300 RA, Leiden, The Netherlands;\\
}

\date{Accepted XXX. Received YYY; in original form ZZZ}

\pubyear{2023}

\begin{document}
\label{firstpage}
\pagerange{\pageref{firstpage}--\pageref{lastpage}}
\maketitle

\begin{abstract}
Hot Jupiters, particularly those with temperature higher than 2000\,K are the best sample of planets that allow in-depth characterization of their atmospheres. We present here a thermal emission study of the ultra hot Jupiter WASP\mbox{-}103\,b observed in two secondary eclipses with CFHT/WIRCam in $J$ and $K_{\rm s}$ bands. By means of high precision differential photometry, we determine eclipse depths in $J$ and $K_{\rm s}$ to an accuracy of 220 and 270\,ppm, which are combined with the published HST/WFC3 and Spitzer data to retrieve a joint constraints on the properties of WASP-103\,b dayside atmosphere. We find that the atmosphere is best fit with a thermal inversion layer included. The equilibrium chemistry retrieval indicates an enhanced C/O (1.35$^{+0.14}_{-0.17}$) and a super metallicity with [Fe/H]$=2.19^{+0.51}_{-0.63}$ composition. Given the near-solar metallicity of WASP-103 of [Fe/H]=0.06, this planet seems to be $\sim$100 more abundant than its host star. The free chemistry retrieval analysis yields a large abundance of FeH, H$^{-}$, CO$_2$ and CH$_4$. Additional data of better accuracy from future observations of JWST should provide better constraint of the atmospheric properties of WASP-103b.
\end{abstract}




\section{Introduction}

The recent decades have seen vast progress in the study of the atmospheres of extrasolar planets, thanks to the rapid developments of both observation technique and planet spectral modelling. Transiting exoplanets, especially hot Jupiters (HJs) and ultra-hot Jupiters (UHJs), are key objects that provide a wealth of vital information about their atmospheres as well as their systems \citep{2010ARA&A..48..631S}. HJs are gas giant planets that have very short orbital periods with P$<$10\,days, and probably tidally locked to their parent stars, which have dayside equilibrium temperature $T_{\rm eq}>1000$K \citep{2018ARA&A..56..175D}. \ww{UHJs have shorter orbital periods and receive irradiation $10-100$ times the insolation of classice HJs, resulting in the equilibrium temperatures in excess of 2000\,K \citep{baxter2020transition}. Under such conditions, most molecules including water become partially thermally dissociated and alkalies are mostly ionized \citep{fortney2008unified}, leading to one of the most significant differences between HJs and UHJs, i.e., water is common in HJ but is missing in UHJs. Besides, simulations indicate that the dayside hemispheres of UHJs have temperature inversion layers because the absorption of the stellar radiation by species such as metals, metal hydride and metal oxides is strong \citep{2019ApJ...876...69L, 2019MNRAS.485.5817G, 2022A&A...659A...7Y}. Therefore, HJs and UHJs are expected to be different in their T-P profiles and chemistry regime, and their comparisons should shed constraint on the atmosphere modelling of exoplanets.} 

Both HJs and UHJs are the most preferable targets for observational studies of planet atmospheres, given their large planet-to-star radius ratio, high $T_{\rm eq}$ and inflated atmospheres.  Observing them at multiple wavelengths allow us to measure the spectral features arising from planetary atmospheres, with which we are able to constrain the temperatures, chemical abundances and mixing ratios of the atmospheres \citep{2009Natur.462..301D, 2016SSRv..205..285M}. Such studies are mainly carried out based on transmission and emission spectra obtained by spectrophotometric or multi band photometric observations during primary transits and occultations (or secondary eclipses) of HJs.

Transmission spectra are known to preferably probe high-altitude atmospheres and outer atmospheres at the terminator regions. Some planets have clear or not very cloudy atmospheres that allow detection of molecules and atoms including H$_2$O, alkali metals Na and K \citep[e.g.][]{2013ApJ...774...95D, 2014MNRAS.437...46N, 2017A&A...608A.135C}, while others may bear high-altitudes clouds/haze and may be too opaque for any spectral features to be detected \citep[e.g.][]{2013MNRAS.436.2974G, 2017A&A...608A..26M, 2016Natur.529...59S}.

On the other hand, thermal emission spectra are sensitive to the atmospheric temperatures at a range of pressures and heights, and thus can in principal provide constraints on the vertical thermal and chemical profiles of the dayside atmospheres. Such observations have been done on several dozens of planets in the last decade using a wide range of facilities. Theoretical interpretations to these data have led to the discoveries of thermal inversions \citep[e.g.][]{2008ApJ...678.1436B, 2009AAS...21420106K, 2018AJ....156...17K}, non-equilibrium chemistry \citep[e.g.][]{2010Natur.464.1161S, 2011ApJ...729...41M} and carbon-rich atmospheres \citep[e.g.][]{2011ApJ...743..191M, 2011Natur.469...64M}. Among them, the \textit{Hubble Space Telescope}(HST) plays a crucial role, particularly for the study of water content using the HST's Wide Field Camera 3 (WFC3) G141 grism. Thanks to the high quality spectra obtained with HST, population studies have been performed recently, resulting in major advance towards the understanding of the atmospheres of HJs \citep[c.f.][]{2016Natur.529...59S, 2021NatAs...5.1224M}. \ww{Analysing transmission spectra of 19 exoplanets spans from cool mini-Neptunes to hot Jupiters, \cite{2019ApJ...887L..20W} conducted a homogeneous survey of Na, K, and H$_2$O abundances and reported a mass-metallicity trend of increasing H$_2$O abundances with decreasing mass, which provided new constraints on the formation mechanisms of the gas giants. \citet{2022arXiv221100649E} suggested that the general lack of trends seen in the transit spectroscopy of the gaseous exoplanets (including WASP-103b) could be either due to the insufficient spectral coverage, or the lack of a simple trend across the whole population or the essentially random nature of the target selection. Combining the data of HST and Spitzer, \citet{2022ApJS..260....3C} presented an retrieval analysis of 25 HJs and UHJs (including WASP-103b) on emission spectra and claimed an apparent link between the abundance of optical absorbers in their atmospheres and the temperature structures.}

Nevertheless, the HST/WFC3 data covers the wavelength range from 1.1$\,\micron$ to 1.7\,\micron, \ww{where H$_2$O normally have strong spectral features, while other prominent molecules like CO, CO$_2$ are relatively weak}. 
The knowledge of CO and CO$_2$ may help to constrain the atmospheric C/O ratio, which is connected to the question where these planets have formed, when comparing with the C/O ratio of their parent stars. Thus, measurements of thermal emission in other bands are vital for a more completed picture of planet atmospheres, especially when combined with HST NIR data. In addition, extending wavelength coverage, particularly to longer wavelengths, would help to place stringent constraints on the atmospheric thermal profiles of the targeted planets and to detect possible thermal inversion layers \citep[e.g.][]{2011ApJ...743..191M}. 

WASP-103\,b is an UHJ with a mass of $\sim 1.5 M_{\rm Jup} $, an inflated radius of $\sim 1.5 R_{\rm Jup}$. It orbits a relatively bright ($V= 12.1$, $K = 10.8$) F8V main-sequence star with a short period of $\sim 22.2$\,hrs \citep{2014A&A...562L...3G}. As a probably tidally locked planet, WASP-103\,b may possess an extremely temperature gradient between its two hemispheres. Its brightness temperature on the dayside is estimated to be $\sim 2500-3200$\,K, as given by the Spitzer measurements of eclipse depths at 3.6\micron\,and 4.5\micron\,\citep{2018AJ....156...17K, 2019MNRAS.489..941P}. A nearby fainter star at an angular separation of 0.242$\pm0.016\arcsec$ to WASP-103 is detected, which may lead to flux contamination to the unresolved photometric observations \citep{2016MNRAS.463...37S, 2016ApJ...827....8N}.

\citet{2017A&A...606A..18L} detected enhanced absorption of alkali (Na, K) in the transmission spectrum (550-960\micron) of WASP-103b using Gemini/GMOS, however they could not confirm the previously-inferred anomalous slope at bluer optical wavelengths \citep{2015MNRAS.447..711S, 2016MNRAS.463...37S}. Later, \citet{2020MNRAS.497.5155W} presented ground-based VLT/FORS2 observations revealing a featureless transmission spectrum (400-600\micron) with neither alkali metal absorption nor strong scattering slope. However, combined with the additional data from Gemini/GMOS, HST/WFC3 and Spitzer observations, H$_2$O absorption was detected at 4.0$\sigma$ level, suggesting a relatively clear atmosphere in the terminator region. Very recently, \citet{2021AJ....162...34K} presented a thorough study based on their new ground-based optical transmission spectrum and archival data of WASP-103b. \ww{They detected a downwards slope at bluer wavelengths that was best fit by unocculted faculae, and claimed weak detection of H$_2$O, HCN and TiO.}

\citet{2017AJ....153...34C} using HST/WFC3 obtained a featureless emission spectrum in near-IR, which suggested the presence of thermal inversion layer or cloud/haze which had muted spectral signals from molecules and atoms. Based on the data obtained by HST/WFC3 and \textit{Spitzer}/IRAC, \citet{2018AJ....156...17K} concluded that the phase-resolved spectra in the WFC3 bandpass are consistent with blackbody emission, while that in the \textit{Spitzer} bands seems to have higher brightness temperature, which is likely caused by CO emission and thermal inversion. \ww{Water vapor was recently detected in WASP-103b by \citet{changeat2022spectroscopic} with depleted abundances of $\sim 10^{-5}$, consistent with the thermal dissociation of this molecule.}
Re-analysed the same HST and \textit{Spitzer} data, \citet{changeat2022spectroscopic} confirmed the existence of the day-side thermal inversion and attributed it to the thermal dissociation of H$_2$O, based on the 1D and 1.5D retrievals of the full-phase observations of WASP-103\,b. In addition, they determined a solar metallicity with log $Z= -0.23^{+0.38}_{-0.44}$ and a solar C/O ratio of $0.57^{+0.17}_{-0.23}$.

In this paper, we present a thermal emission study based on our ground-based high-precision eclipse observations of the hot Jupiter WASP-103\,b in $J$ and $K_{\rm s}$ band with CFHT in 2015 (PI: Wei Wang), and an archival CFHT $K_{\rm s}$ band data taken in 2014 (PI. Delrez). Details of the observations are described in section~\ref{sec:obs}, following by introduction of data reduction and analysis in Section~\ref{sec:anal}. We compare the observations to model spectra in Section~\ref{sec:Retrieval Analysis} and summary our discussion and conclusion in the final section~\ref{sec:summary}.

\section{Observations and data reduction}
\label{sec:obs}

\begin{figure}
   \centering
  \includegraphics[width=8cm, angle=0]{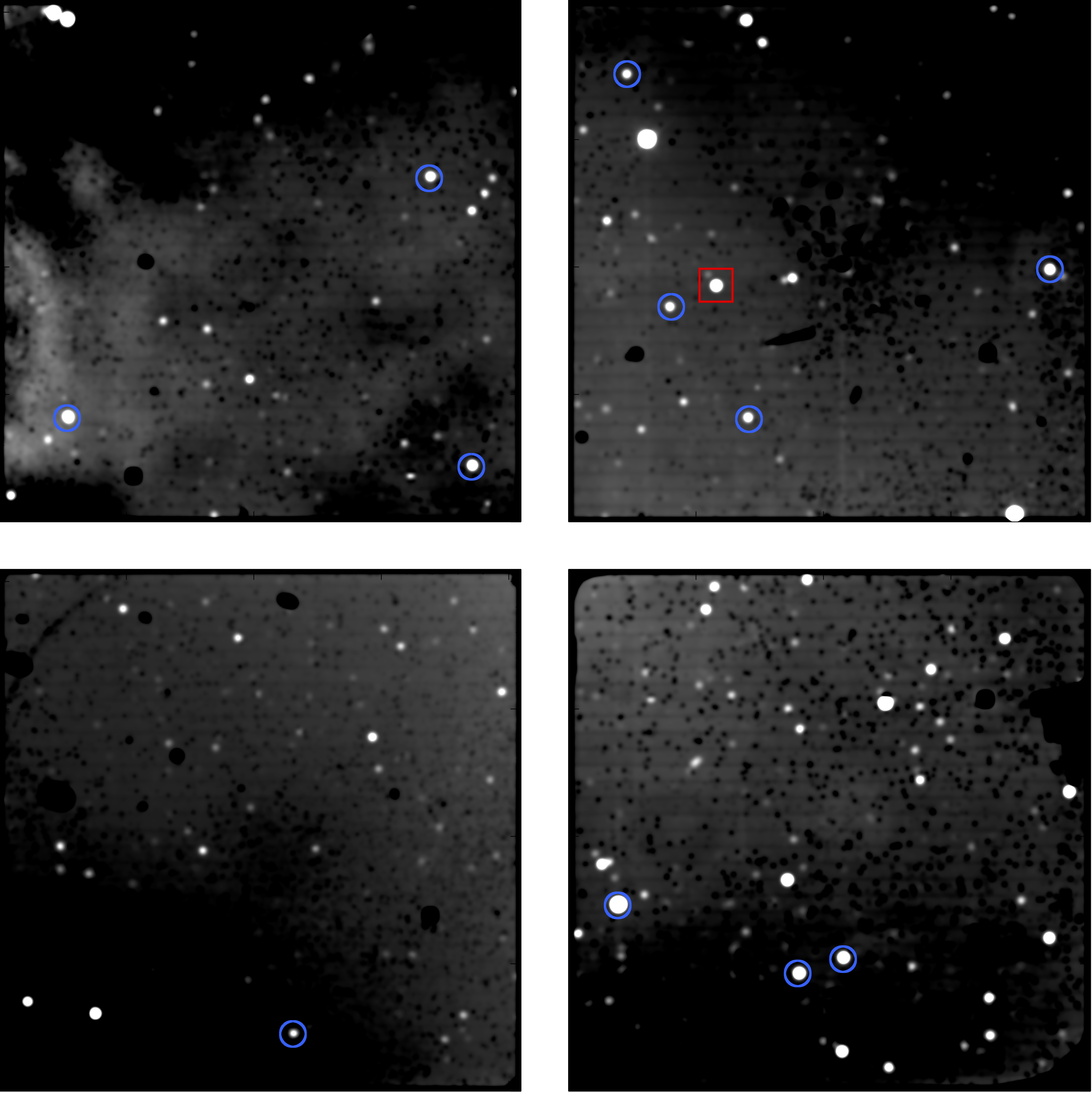}
   \caption{CFHT/WIRCam full frame reduced image during our observation of WASP-103\,b in  the $J$ band. WASP-103b is marked with a red square and the final eleven reference stars that were used are marked as blue circles.} 
   \label{Fig1}
\end{figure}

We observed two secondary eclipses of WASP-103\,b with the Widefield InfraRed Camera (WIRCam; \citet{2004SPIE.5492..978P}) on the Canada-France-Hawaii Telescope (CFHT). WIRCam has a large field of view (FOV) of 21\arcmin$\times$21\arcmin, which allows to cover a large number of appropriate reference stars for differential photometry. The two observations were taken respectively in the $J$ band on 2015 June 9, and in the $K_{\rm s}$ band on 2015 May 28. The total observation time for each run is $\sim$5.5 hours, covering the entire eclipse assuming a circular orbit, and $\sim$2.5 hours out-of-eclipse observations. The airmass ranges from 1.02$-$1.37, and 1.02$-$2.01 for the $J$- and $K$s band respectively. The night of June 9 was photometric with a median seeing of 0.88\arcsec, however the May 28 night was not ideal bearing large sky variations. The atmospheric extinctions of two observations are reported to be $\sim$0.06\,mag. \ww{As described in Section 3.3, we also employed the $K_{\rm s}$-band data taken on 2014 May 20 from 06:50 to 12:40 UT (Program ID: 14AO38, PI: L. Delrez) with a very similar observation strategy. For more details, please refer to Section~\ref{sec:anal lc} in \citet{delrez2018high}.} 

In order to achieve ultrahigh photometric precision down to $\sim10^{-4}$ as required by our science goals, the `Staring Mode' \citep{2010SPIE.7737E..2DD} was adopted in both observations, following previous successful experiments \citep[e.g.][]{2010ApJ...717.1084C, 2010ApJ...718..920C, 2013ApJ...770...70W}. The Staring Mode is designed to keep the pointing of the telescope as stable as possible, so that the effect of intra-pixel and inter-pixel variations of detectors on photometric precision could be largely weakened.  

The full WIRCam array is used to include reference stars as many as possible. The telescope was defocused to 1.7\,mm and 1.8\,mm for $J$- and $K_{\rm s}$ band observations, resulting in a donut-shaped PSF with a radius of $\sim$4\arcsec. This is to further minimize the impact of flat fielding uncertainties and intra-pixel sensitivity variations, as well as to keep the flux of the target and reference stars well below the detector saturation level with reasonably long exposure time and thus low fraction of overhead. We applied a detector integration time (DIT) of 10\,s and 15\,s for $J$ and $K_{\rm s}$ bands, respectively, and NDIT=12. The overall duty is 39.8\% for $K_{\rm s}$ band run, and is 50.3\% for $J$ band run. 

Fig.~\ref{Fig1} is a full frame reduced image showing the observing field-of-view (FOV), with the target star and the final selected reference stars marked as black squares and red circles, respectively. Given that WASP-103 is bright, and the exposure times and defocus amounts that were set for our run for the purpose of minimizing systematics, the number of available reference stars in the FOV is limited. This results in relatively large observation uncertainties as compared to theoretical predictions from \citet{2010SPIE.7737E..2DD}.

The raw data were reduced using the 'I'iwi pipeline version 2.1.200~\footnote{https://www.cfht.hawaii.edu/Instruments/Imaging/WIRCam/IiwiVersion2Doc.html}, including: flagging the saturated pixels, non-linearity correction, reference pixels and dark subtraction, flat fielding, bad pixels and guide window masking. \ww{The 'I'iwi pipeline construct bad pixel masks from darks and dome flats. The bad pixel mask has 1 for good pixels and the float Not-a-Number (NaN) value for bad pixels.} There might be bad pixels within the photometry aperture of the target and reference stars. In order to improve the accuracy and the signal-to-noise ratio, these bad pixels were flagged, and their values were replaced by interpolating the counts of their adjacent pixels. Aperture photometry was then performed on the reduced image for the target star and dozens of candidate reference stars that were not used as guiding stars. Fractional contribution of pixels at the edge of the circular aperture were taken into account. Stars of similar brightness with ($\Delta$m$<2.5$ mag) in the FOV, particularly those in the same WIRCam chip, were selected as candidate reference stars. We kicked out the reference stars which whose out-of-eclipse light curves are quite different from others, or possess relatively large photometric variation. For each star, 11 different aperture radii were applied with a step of 0.5 pixels. The final aperture radii and the final selection of reference stars will be determined in the iteration procedures of finding best-fit light-curves, as described in details in the following sections.

\section{Data analysis and results}
\label{sec:anal}

\subsection{Light curve derivation and modelling}
\label{sec:anal lc}

We start from the technique presented by \citet{2001PASP..113.1428E} to obtain high precision differential light curves with various aperture sizes and different combinations of reference stars. Firstly, for each aperture diameter $D$ and each reference star, each raw light curve is normalized individually with their median values determined by $2-3\sigma$ clipping algorithm. Then, an average reference light curve at a $D$ and a reference star group (RSG), $L_{\rm ref(D, RSG)}$, is obtained by taking the weighted averages of each normalized light curve, with the weights being the inverse of photometric errors. Then, the average reference light curve is employed to differentiate the target light curve to remove systematics. A 5 $\sigma$ cut is applied to remove obvious outliers in the normalized target light curve. Besides, there is an obvious time-correlated trend in the normalized observed light curve, which is referred as the background trend and could be roughly modelled as a linear function, defined by the out-of-eclipse light curve with time:
\begin{equation}
{ B{_f}=c{_1}+c{_2}t} ,
\end{equation}
where $c_1$, $c_2$ are fitting parameters, $t$ is time relative to the middle eclipse time in the unit of seconds. 

\begin{figure}
   \centering
  \includegraphics[width=8.5cm, angle=0]{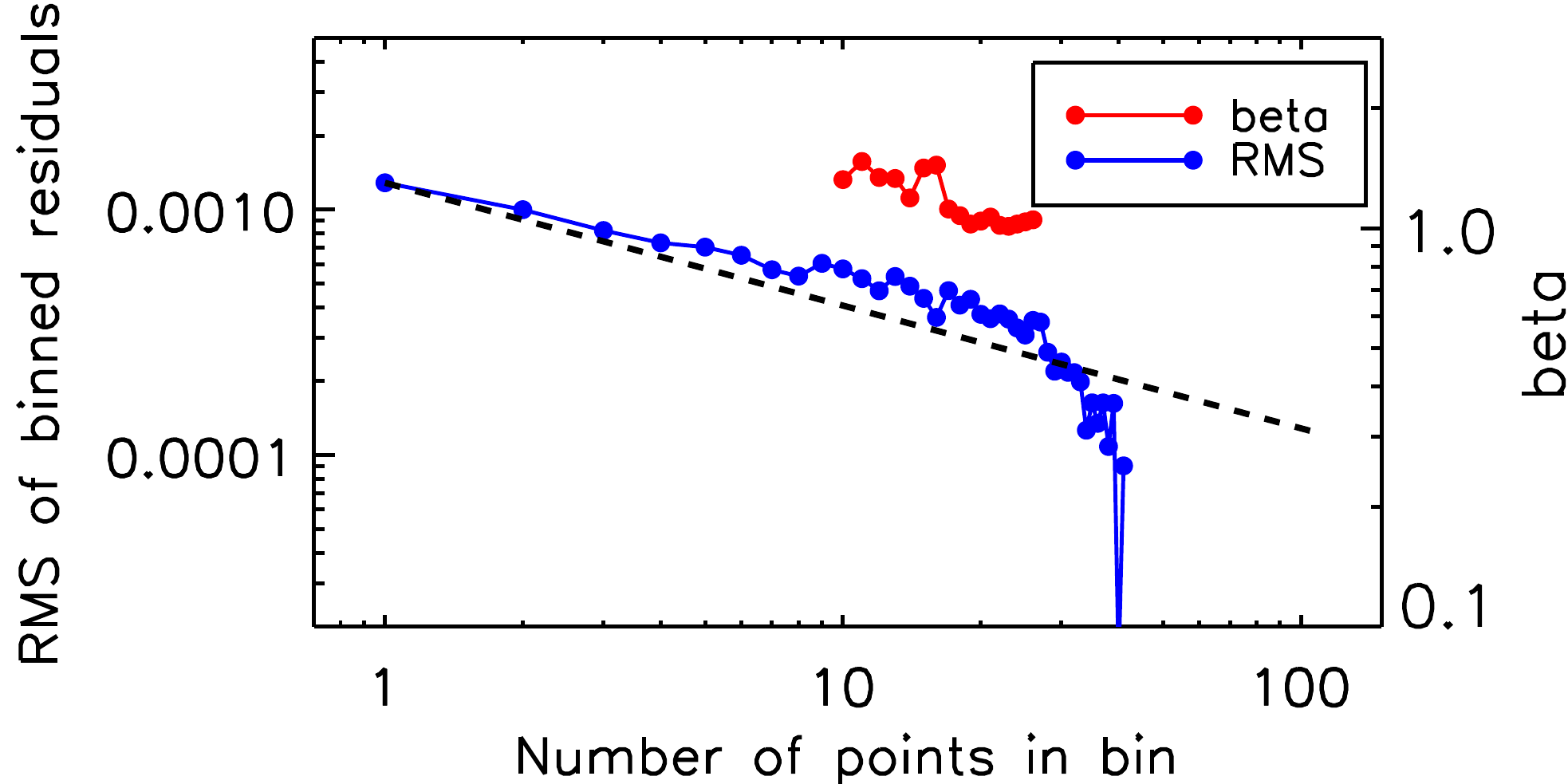}
   \caption{RMS and noise factor $\beta$ of our residuals to the best-fit model for the various data sets in $J$ band. } 
   \label{Fig:rms-beta}
\end{figure}

The target light curves are then modelled using the IDL package {\tt\string EXOFAST}\footnote{https://exoplanetarchive.ipac.caltech.edu/docs/exofast/exofast\_home.html} \citep{2013PASP..125...83E}. This suite of routine uses Markov Chain Monte Carlo (MCMC) methods to fit the model to \citet{2002ApJ...580L.171M} observations \citep{2005AJ....129.1706F}. Background baseline trend has been observed in almost all previous NIR eclipse light curves \citep{2010ApJ...717.1084C, 2010ApJ...718..920C, 2013ApJ...770...70W, delrez2018high}, which probably originate from the variations in instruments and/or atmospheric conditions \ww{and should be removed or modelled to reveal the planet signal. In this work, $B_{f}$ is firstly detrended with a linear function of $t$ to the out-of-eclipse part light curve. The remaining baseline is then fit jointly with a polynomial function and the eclipse model, to avoid possible overfitting using \texttt{EXOFAST}. The polynomial function we used is dependent on the time relative to the eclipse center in seconds $t$, the image coordinates of the target star $x$ and $y$ in pixels, and the width $w$ in pixels of the target's point spread function (PSF), as described in Formula~2.}

To evaluate the amount of correlated noise in each light curve $L_{\rm ref(D, RSG)}$, the best-fit model$-$data residuals are binned down to be compared with the expected Gaussian noise of one over the square root of the bin size (Fig.~\ref{Fig:rms-beta}). To quantify the amount of correlated noise in our data sets, the parameter $\beta$ defined by \citet{2008ApJ...683.1076W} is used, which is the ratio of the residuals to the Gaussian noise expectation (see Fig.~\ref{Fig:rms-beta}). To determine $\beta$, we calculate the average ratios with bin sizes of 15 to 85 data points, as shown in red in Fig.~\ref{Fig:rms-beta}. It sometimes occurs oddly that the scaled-down residuals are smaller than the Gaussian noise expectation, i.e., $\beta<1$. For such cases, we set $\beta=1$.

\subsection{Optimal techniques to determine the ``best'' light curves}
\label{sec:anal bestlc}
It is realized that for almost each combination of $D$ and RSG, an eclipse-like light curve could be found with different measured eclipse depths, errors, and best-fit residuals. This fact, on one hand, assures that our data do detect thermal emission from the target planet. While on the other hand, this calls the requirement for a further step to find the ``best'' combination of $D$ and RSG, which produce the ``best'' light curve that could be fitted by an eclipse light curve model with the smallest residuals, while no hint for overfitting is noticeable. 

For the case of $D$ small but still large enough to contain most stellar photons, the rms of residuals of the light curve with best-fit model subtracted is small, as the impact of sky background is mitigated at small $D$. While for large $D$, the vast majority of the light from stars are included in aperture photometry even during moments of poor seeing and guiding, thus red noise could be best fit and removed, resulting $\beta \sim 1$. Therefore, we use rms and $\beta$ as proxies to achieve a balance between the two competing factors: avoiding high sky background that arises from with large $D$, and reducing the presence of time-correlated noise that comes along with small $D$. In this work, we explore all the combinations RSG and various values of $D$, to seek the minimum value of the metric rms$\times \beta^{2}$, following by the prescription of \citet{2015ApJ...802...28C}. Fig.~\ref{Fig:J_rxb2} shows an example on how to determine the best combination of $D$ and RSG.

\subsection{Light curve fitting}
\label{sec:anal lcfitting}

\begin{figure}
   \centering
  \includegraphics[width=8cm, angle=0]{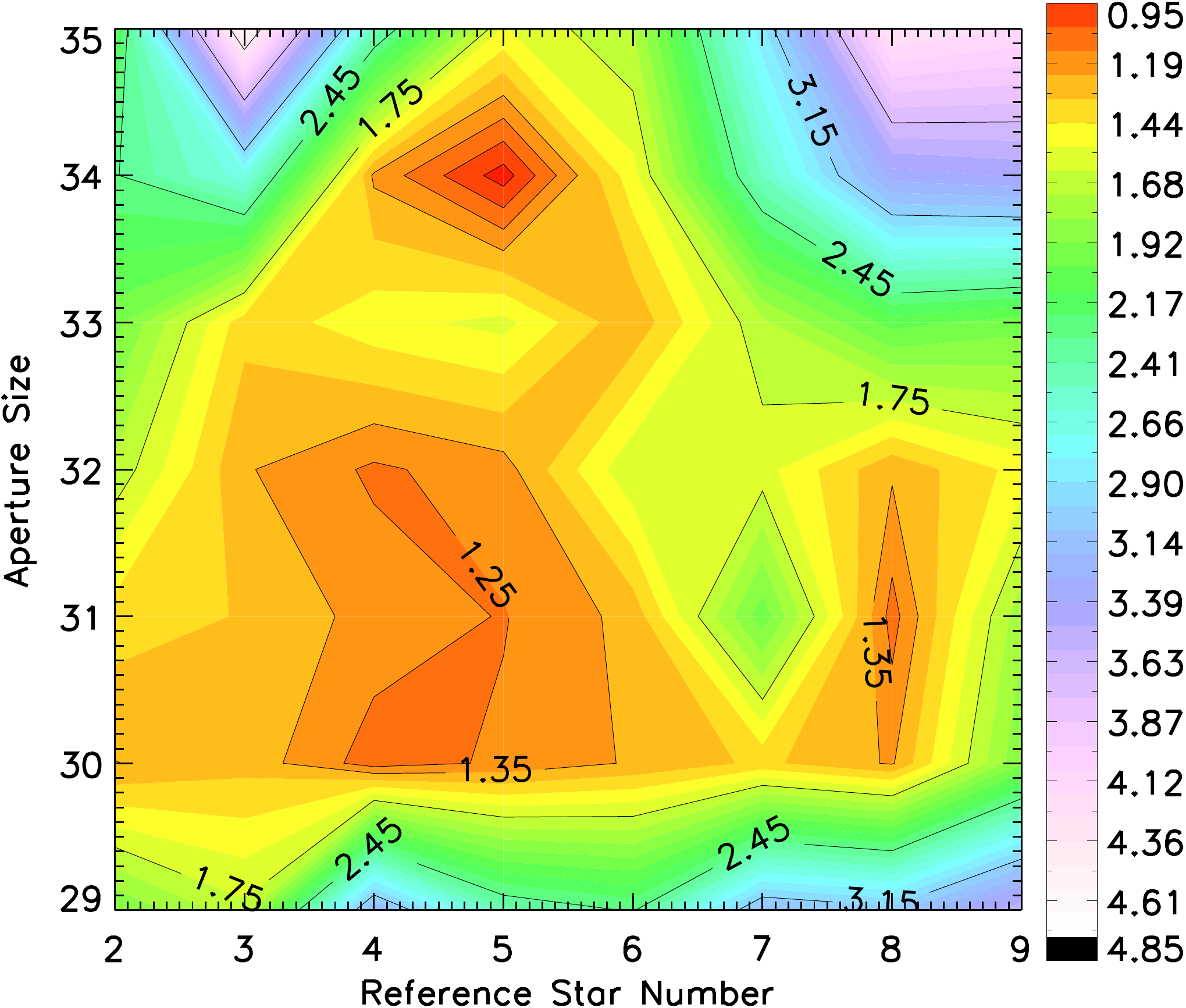}
   \caption{The normalized rms$\times\beta^{2}$ distribution overlaid with contour maps for the various combinations of aperture diameter $D$ and the number of reference star group ($N_{\rm RSG}$) in the $J$ band. The minimum value of rms$\times\beta^{2}$ is reached with $D=34$ and $N_{\rm RSG}=5$.} 
   \label{Fig:J_rxb2}
\end{figure}

\begin{figure}
   \centering
  \includegraphics[width=8cm, angle=0]{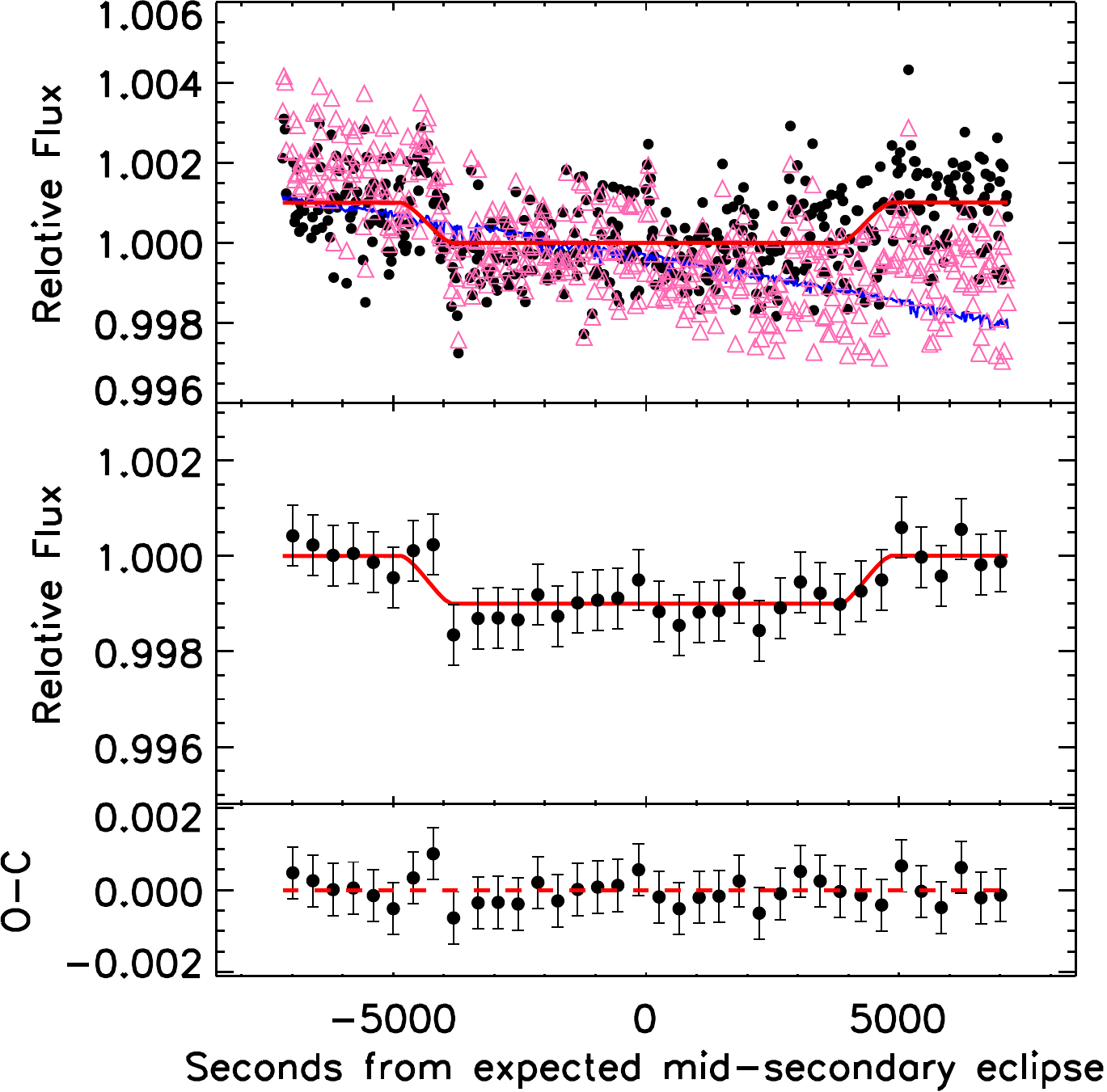}
   \caption{The secondary eclipse of WASP-103b observed by CFHT/WIRCam in $J$ band on 2015 June 9. The top panel shows the unbinned light curve with the polynomial function fitting of the red noise (blue line) and the unbinned light curve subtracted red noise with the best-fit secondary eclipse model (red line). The second panel shows the binned light curve divided by the best-fit model (red line). Bottom panel shows the binned residuals from the best-fit model. The RMS of the binned residuals is 654 ppm~(7.8min bins). The contamination from the nearby star are not corrected here.}
   \label{Fig:J_LC}
\end{figure}

Following the methods described above, light curves $L_{\rm ref(D, RSG)}$ are obtained for various combinations of aperture size $D$ and RSG, and the best combination of $D$ and RSG is found by finding the minimum value of rms$\times \beta^{2}$. As shown in Fig.~\ref{Fig:J_rxb2}, for $J$ band data, choosing $D=34$ and including the best 5 reference stars yield the best fitting results. In this step, the same $D$ is used for the target and reference stars. After the best combination of $D$ and RSG is found, we further look for the best light curve by varying individually $D$ of each reference stars with a step of $\pm$1 pixel. The best-fit light curve is shown in Fig.~\ref{Fig:J_LC}, for which an eclipse depth of $1170\pm190$ is derived. The resulting $D$ of each reference star is slightly different from the same-$D$ fitting result. There are about $\sim3-5$ data points exceeding 5 times the standard deviation of the corrected target light curve, which are marked as outliers and are removed from model fitting.

Assuming a circular orbit, we produced the $J$ band secondary eclipse light curve model by setting SECONDARY \& SPECPRIORS keywords of {\tt\string EXOFAST} and used the latest published parameters as priors. For the implementation of the secondary eclipse case, we enforced the limb darkening coefficients to be zero. Our model includes 12 prior parameters in total. A de-correlatively polynomial function was applied in order to fit the baseline curve, with the function:
\begin{equation}
{ f=1+a{_1}x+a{_2}x^2+a{_3}y+a{_4}y^2+a{_5}t+a{_6}t^2+a{_7}r+a{_8}r^2} ,
\end{equation} where a$_{1,2...8}$ are the coefficients to be derive, $x$, $y$ are the positions of the centroid of the target star, $t$ is the time relative to the eclipse center in seconds, and $r$ is the stellar PSF full-width at half maximum (FWHM). We run 10$^{6}$ steps in the MCMC chain, yielding best-fit results as listed in the second column of Table~\ref{tab1} and \ww{the posterior distribution plots are shown in Fig.~\ref{Fig:exofast}.}

\begin{table*}
\centering
\caption[]{Best-fit Secondary Eclipse Parameters in $J$ \& $K_{\rm s}$}\label{tab1}

\begin{tabular}{lcc}
 \hline\noalign{\smallskip}
Parameter & $J$ & $K_{\rm s}$ \\
\noalign{\smallskip}
\hline
\noalign{\smallskip}
Stellar parameters \\
$M_{*}$~($M_{\sun}$)& $1.223_{-0.080}^{+0.085}$ & $1.225_{-0.082}^{+0.085}$ \\
$R_{*}$~($R_{\sun}$)& $1.433_{-0.032}^{+0.033}$ & $1.434\pm0.033$ \\
$L_{*}$~($L_{\sun}$)& $2.59_{-0.40}^{+0.46}$ & $2.58_{-0.41}^{+0.45}$ \\
$\log g_*$~(cgs)& $4.2129_{-0.010}^{+0.0099}$& $4.2131_{-0.010}^{+0.0099}$ \\
$T_{\rm eff}$~(K) & $6110\pm200$ & $6120\pm200$  \\
$[{\rm Fe/H}]$~(dex) & $0.06\pm0.13$& $0.06\pm0.13$ \\
\noalign{\smallskip}
\hline
\noalign{\smallskip}
Planet Parameters\\
$P$~(days)& $0.925541_{-0.000020}^{+0.000019}$& $0.925542\pm0.000019$ \\
$a$~(AU)& $0.01987_{-0.00044}^{+0.00045}$& $0.01988_{-0.00046}^{+0.00045}$ \\
$R_{\rm P}$~($R_{\rm Jup}$)& $1.524_{-0.044}^{+0.045}$& $1.525_{-0.045}^{+0.046}$ \\
$T_{eq}$~(K)& $2505\pm83$& $2504_{-83}^{+82}$ \\
$R_{\rm P}/R_{*}$& $0.1094\pm0.0020$& $0.1094\pm0.0020$ \\
$a/R_{*}$& $2.9822_{-0.0069}^{+0.0068}$& $2.9823\pm0.0067$ \\
\noalign{\smallskip}
\hline
\noalign{\smallskip}
Secondary Eclipse Parameters\\
$T_{\rm mid}$~($BJD_{\rm TDB}$)& $57182.40589_{-0.00022}^{+0.00023}$&$56797.38610\pm0.00023$ \\
$b$~($R_{*}$)& $0.0731_{-0.0061}^{+0.0060}$& $0.0729\pm0.0060$ \\
$T_{12}$~(days)& $0.01150\pm0.00021$&  $0.01150\pm0.00021$\\
$T_{14}$~(days)& $0.11207\pm0.00035$& $0.11208\pm0.00035$ \\
$F_{\rm p}/F_{\ast}^{a}$ & $0.00117\pm0.00019$ & $0.00113\pm0.00023$ \\

  \hline\noalign{\smallskip}
  \noalign{\smallskip}
\end{tabular}

$^a$: the derived eclipse depths listed here are not yet corrected for the dilution by the nearby star.
\end{table*}

\begin{figure}
   \centering
  \includegraphics[width=8cm, angle=0]{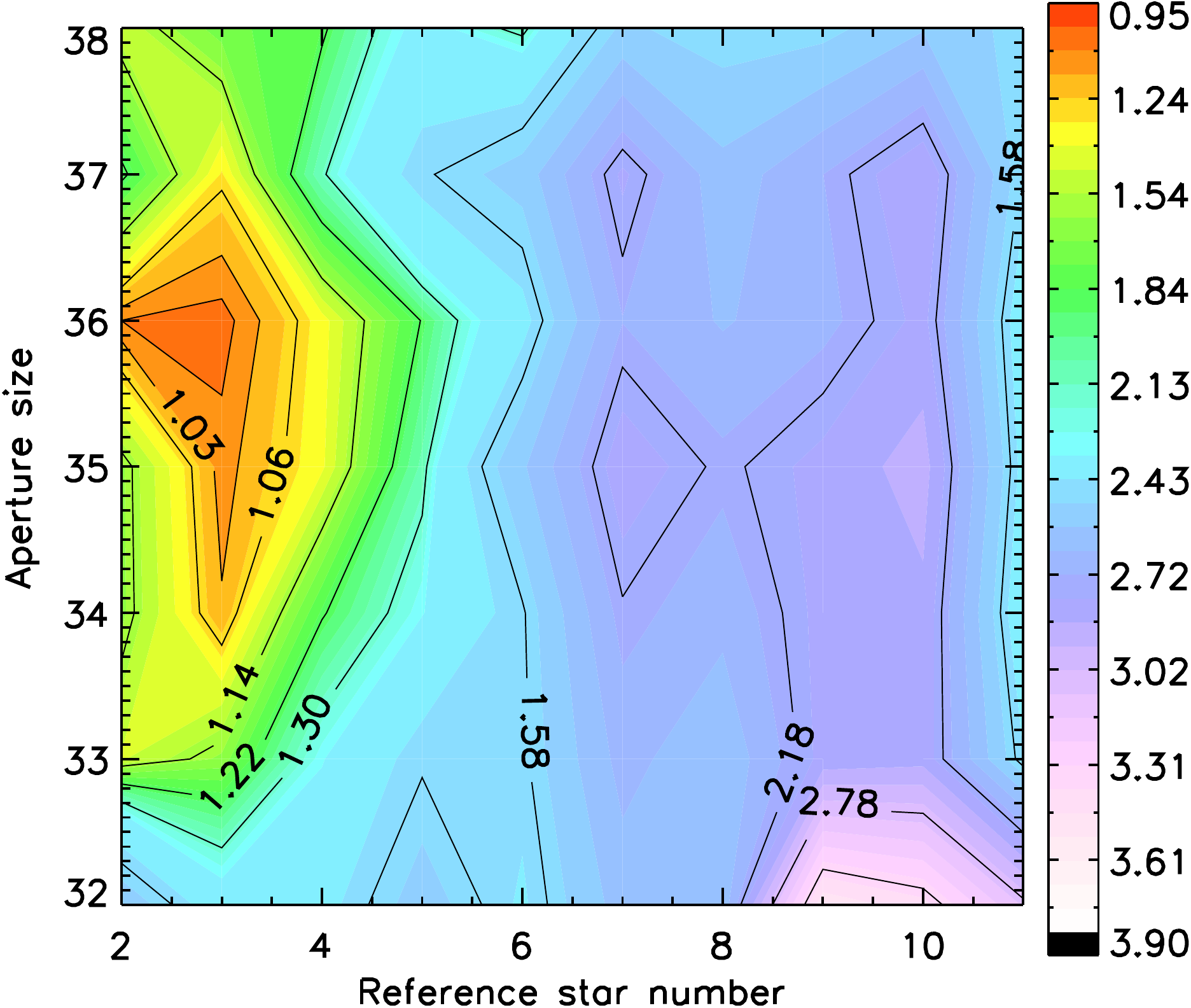}
   \caption{Same as Fig.~\ref{Fig:J_rxb2} but for the $K_{\rm s}$ band. The minimum value of rms$\times\beta^{2}$ is reached at $D=36$ and $N_{\rm RSG}=3$.} 
   \label{Fig:Ks_rxb2}
\end{figure}

\begin{figure}
   \centering
  \includegraphics[width=8cm, angle=0]{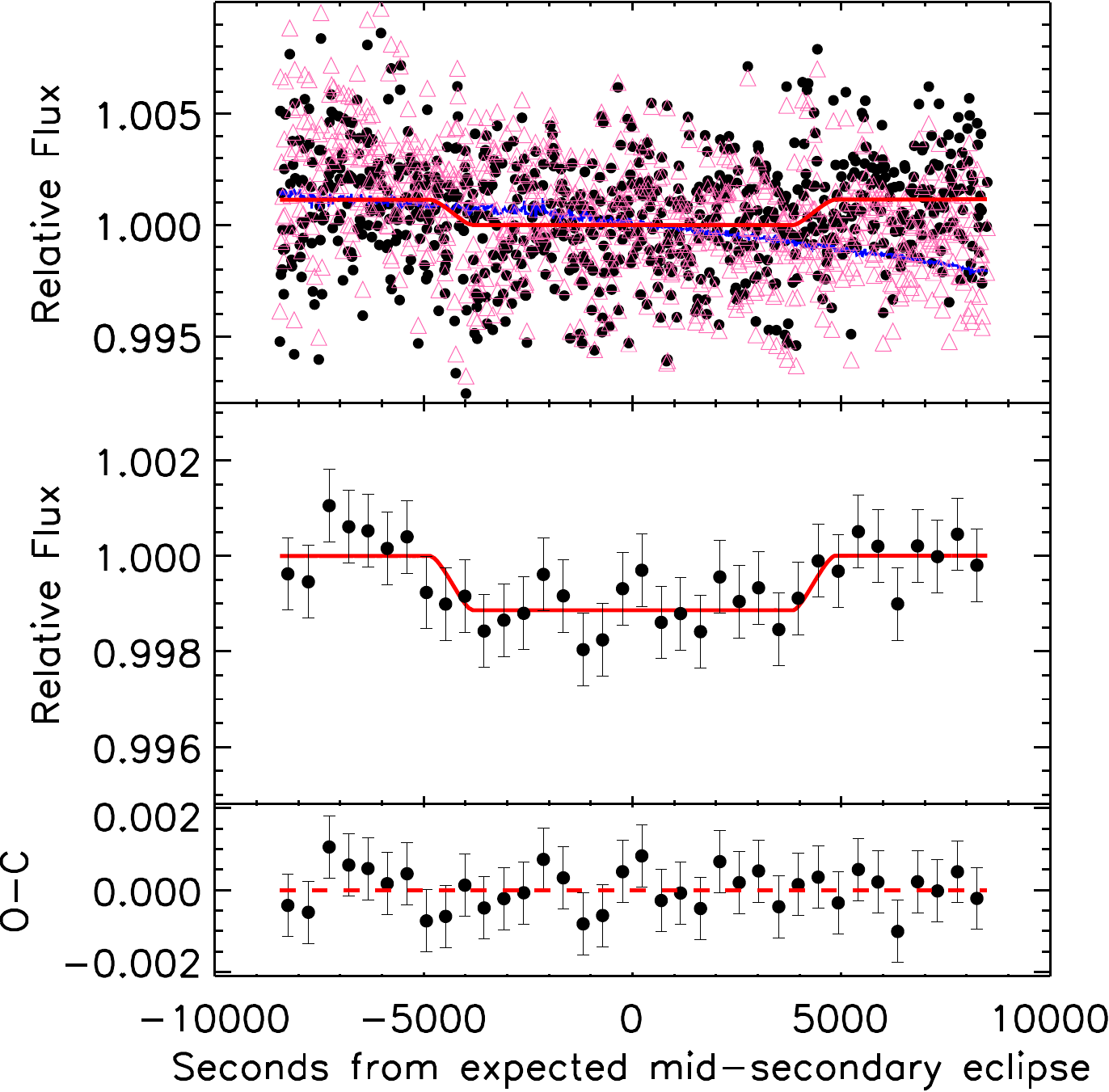}
   \caption{Secondary eclipse light curve of WASP-103\,b in $K_{\rm s}$ band. Same as Figure.~\ref{Fig:J_LC}. The binned residual RMS is 775\,ppm(8min bins).} 
   \label{Fig:Ks_LC}
\end{figure}

Due to the bad observable conditions in 2015 May 28, the data set obtained for $K_{\rm s}$ band does not meet the precision requirement for our scientific goal. Therefore, we employed the data taken on 2014 May 20 from 06:50 to 12:40 UT (Program ID: 14AO38, PI: L. Delrez) with a very similar observation strategy, to determine the $K_{\rm s}$ eclipse depths. For details of this data, please refer to \citet{delrez2018high}. We note that the 2014 $K_{\rm s}$ band observation has a FOV  slightly different from that of the $J$ band observation, thus a new RSG and $D$ are to be found. The data then were reduced and analyzed in the same manner as described above for the $J$ band data. As a result, shown in Fig.~\ref{Fig:Ks_rxb2}, choosing $D_{\rm ref_{i}}=36$ and 3 reference stars yield an initial best-fitting result. Similarly, we applied a 3$\sigma$ threshold to remove the outlying data points to obtain a ``cleaner'' target light curve, with which we obtained the best differential light curve in $K_{\rm s}$, as presented in Fig.~\ref{Fig:Ks_LC}. The derived  parameters are listed in the third column of Table~\ref{tab1}. 

We note that the data points at the beginning of each observation have a relatively large systematic error, therefore we only used the data between with $T-T_{\rm mid}$ between $-8500$\,s and 8500\,s when performing light curve fitting and eclipse depth determination. It is also notable that our measurement of the $K_{\rm s}$ band secondary eclipse depth is quite different ($>5\sigma$) from those reported by \citet{delrez2018high}, which might be caused by different baseline continuum shapes that may have a different impact on the quadratic terms of baseline fitting. Note that their derived eclipse depth is higher than their model predictions (c.f. Fig.~5, in their paper).

\subsection{Flux Decontamination and Eclipse depths}
\label{subsec:eclipse depth}
The ``measured'' eclipse depth from the unresolved photometric light curve may be underestimated if a nearby companion is included in the aperture. Therefore, flux contamination from nearby companions must be corrected before determining the planet-to-star flux ratio $F_{\rm P}/F_{\rm S}$. WASP-103 was reported to have a close companion, that may dilute the measurements of its eclipse depths. \citet{2015A&A...579A.129W} found that the companion is $0.242\pm$0.016\arcsec from their $i$ and $z$ band observations. \citet{2016ApJ...827....8N} provided further evidence in $JHKs$ with the measured magnitudes difference of $\Delta J$ = 2.427 $\pm$ 0.030, $\Delta H$ = 2.2165 $\pm$ 0.0098, $\Delta Ks$ = 1.965 $\pm$ 0.019. They claimed that the companion might be bounded to WASP-103. This conclusion was later confirmed by \citet{2017AJ....153...34C}, but was questioned by a more recent work by \citet{2022A&A...657A..52B}, as indicated by a non-detection of the visual companion via lucky imaging observations and the RV observations with CORALIE. Whether the fainter star is bounded to WASP-103 or not, at such a small angular separation, it would anyway contaminate the flux measurements in both transmission and emission spectra. 

The flux contamination ratio, is simply the ratio of the flux of the contaminant to that of the target star, i.e., $F_{\rm con}/F_{\rm W103}$ in our case. Their values are listed in Table~\ref{tab2}, which can be directly calculated using the observed magnitude differences $\Delta J=2.427\pm0.030$ and $\Delta K_{\rm s}=1.965\pm0.019$ from \citet{2016ApJ...827....8N}. Then, $F_{\rm P}/F_{\rm S}$ is given by

\begin{equation}
\frac{F_{\rm P}}{F_{\rm S}}={\rm d}_{\lambda}\times(1-\frac{F_{\rm con}}{F_{\rm W103}})^{-1}
\end{equation}
, where $d_\lambda$ is the ``measured'' eclipse depth before flux decontamination correction at the central wavelength $\lambda$ of each passband.

\begin{table*}

\caption[]{The ``measured" original eclipse depths, the flux contamination ratios and the decontaminated eclipse depths in $J$ \& $K_{\rm s}$ for WASP-103\,b in this work.\label{tab2}}
 
 \begin{tabular}{cccc}
 \hline\noalign{\smallskip}
Filter &  Uncorrected eclipse Depths  &   $ F_{\rm cont}/F_{\rm W103}$   & Corrected Eclipse Depths  \\
 \hline\noalign{\smallskip}
 $J$        & 0.00117$\pm$0.00019    & 0.1070 $\pm$ 0.0029 & 0.00131$\pm$0.00022  \\
 $K_{\rm s}$ & 0.00113$\pm$0.00023    & 0.1637 $\pm$ 0.0029 & 0.00135$\pm$0.00027  \\
\noalign{\smallskip}\hline
\end{tabular}

\end{table*}

The decontaminated secondary eclipse depth we obtained for $J$ band in this work is $1310\pm220$, well consistent with \citet{cartier2016near} and \citet{2018AJ....156...17K}, which obtained an eclipse depth of  $1256^{+154}_{-155}$\,ppm at $\lambda\sim1.252\micron$ and $1480\pm$46\,ppm at $\lambda\sim1.275\micron$, respectively. Note that the eclipse depth from \citet{cartier2016near} is $\sim$10\%  lower than that from \citet{2018AJ....156...17K}, which is likely a result that the latter analysis considers the planetary thermal phase variation, at an order of $\sim100$\,ppm. If we take the amplitude of the effect into account, our result for the $J$ band is only $\sim40$ ppm smaller than that of \citet{2018AJ....156...17K}. Applying a simple linear interpolation, we calculate the result for $K_{\rm s}$ band. However, the amplitude-correct results could not fit the model well, thus, we utilize the uncorrected results in atmospheric model fitting.

\section{The Atmospheric properties of WASP\mbox{-}103b}
\label{sec:Retrieval Analysis}
\subsection{Atmospheric Modelling}
\label{modelling}

\begin{figure}
   \centering
   \includegraphics[width=8cm, angle=0]{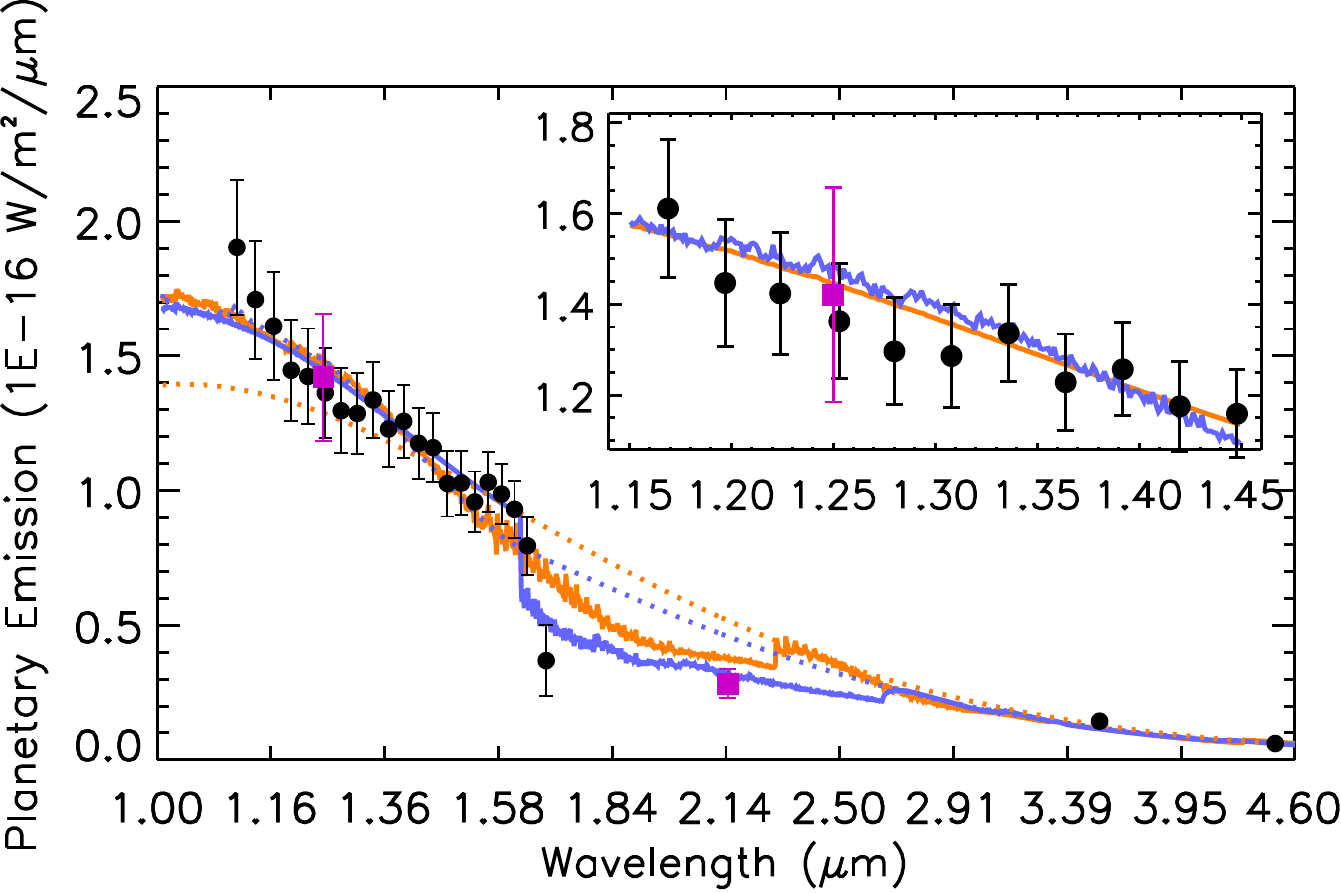}
   \caption{Observations and model spectra of dayside thermal emission of WASP-103b. Our $J$ and $K_{\rm s}$ data are presented by purple solid square, and other archival data are presented by black solid circle. The orange line shows the EQ model, while the violet line shows the FREE model. \ww{The two dotted lines show the EQ and FREE models assuming isothermal. }}
   \label{Fig:moddel}
\end{figure}

Fig.~\ref{Fig:moddel} shows the dayside emission spectra of WASP103\,b to be investigated in this work, including the photometric eclipse depths derived from the ground-based $J$ and $K_{\rm s}$ observations (purple square), those from Spitzer observations~\citep{2018AJ....156...17K}, and those from HST~\citep[black filled circles ][]{cartier2016near}. As reported by several previous work~\citep{cartier2016near, 2018AJ....156...17K}, WASP-103b seems to be featureless in the HST/WFC3 band\ww{, except for the very recent work by \citet{changeat2022spectroscopic}, which reported detection of H-, H$_2$O, CO and CH$_4$.} Therefore adding $K_{\rm s}$ band data may provide crucial information to constrain the planet's atmosphere, particularly the C abundance and the carbon-to-oxygen (C/O) ratio. It is quite interesting the last HST data point and our $K_{\rm s}$ data are significantly lower than the $J$ band data and the other HST data points, which should be related to the opacity of CO$_2$ and/or CO molecules. 

We follow the exoplanetary atmospheric modeling and retrieval technique of \citet{madhusudhan2009temperature} to investigate the dayside atmosphere of this planet. The spectral retrieval analyses are performed using the open-source package \texttt{petitRADTRANS}~\citep{Molliere2019A&A}. We assume a 1D parallel atmosphere with 100 layers, divided equally in logarithmic space from 10$^3$\,Pa to 10$^{-6}$\,Pa, and the \texttt{PyMultiNest}~\citep{Buchner2014} code to calculate the Bayesian evidence, which represent the average likelihood under the prior for a specific model choice. \texttt{PyMultiNest} implements the multimodal nested sampling algorithm based on the MultiNest library~\citep{Feroz2009}.

\begin{figure}
   \centering
  \includegraphics[width=7cm, angle=0]{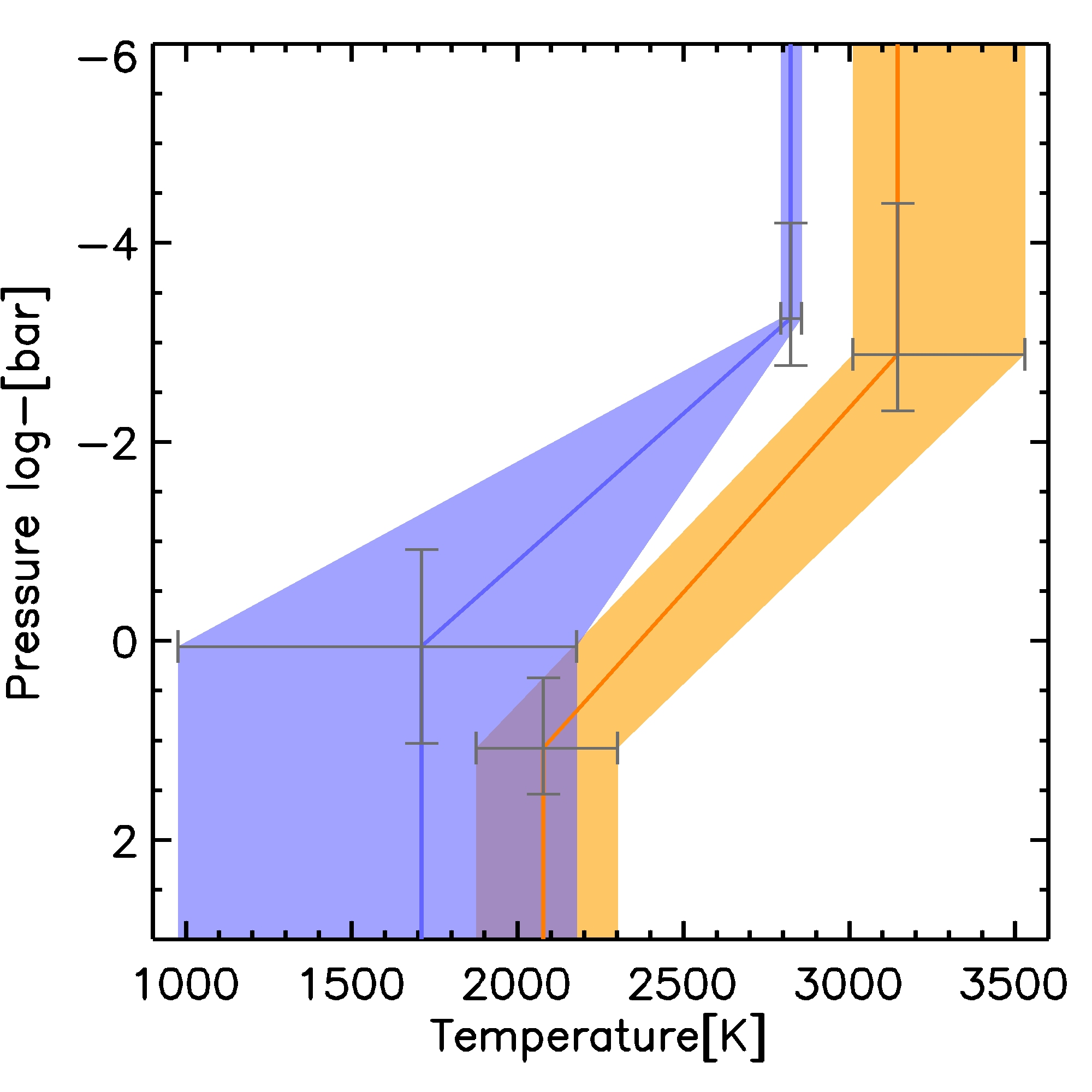}
   \caption{The retrieved vertical T-P profiles for the EQ model (orange) and FREE model (violet) \ww{with 1$-\sigma$ confidence intervals}, respectively.} 
   \label{Fig:TP}
\end{figure}

As pointed out by previous works, WASP-103\,b seems to possess a dayside thermal inversion layer \citep{delrez2018high,2018AJ....156...17K,changeat2022spectroscopic}. Therefore, we adopt a two-point temperature-pressure (T-P) profile \citep{Brogi2014} that allows thermal inversion, which assumes an isothermal atmosphere at altitudes above the lower pressure point ($T_1$, $P_1$) and below the higher pressure point ($T_2$, $P_2$), while $T$ changes linearly with log\,$P$ between the two points. $T$ ranges between 0 and 4000\,K. The retrieved T-P profiles are shown in Fig.~\ref{Fig:TP}, implying the presence of thermal inversion in the atmosphere of WASP-103\,b.

We consider two different models with either free chemistry (hereafter the FREE model) or equilibrium chemistry (hereafter the EQ model). \ww{We are aware that assuming equilibrium chemistry is often oversimplified and may not accurately reflect the actual chemical processes occurring in the atmosphere. Moreover, it can not typically account for the formation of clouds, which could have a significant impact on the temperature and composition of the atmosphere. However, the EQ model is still useful to be served as a starting point for initializing abundances in HJ's atmospheres and for the compaison study with the FREE model results.}

For the EQ model case, only two free parameters, i.e., the C/O number ratio and the metallicity [Fe/H] are to be retrieved. As described in \citet{Molliere2017}, the mass fraction of each species can be interpolated from a chemical table as a function of $P$, $T$, [Fe/H] and C/O. The list of reactant species includes \ww{H$_2$ \citep{2013JQSRT.130....4R}, He \citep{1965PPS....85..227C}, CO \citep{2010JQSRT.111.2139R}, H$_2$O, HCN \citep{2006MNRAS.367..400H}, C$_2$H$_2$, CH$_4$ \citep{2014MNRAS.440.1649Y}, PH$_3$ \citep{2015MNRAS.446.2337S}, CO$_2$, NH$_3$ \citep{2011MNRAS.413.1828Y}, H$_2$S, VO
\citep{Molliere2019A&A}, TiO \citep{2019MNRAS.488.2836M}, Na \citep{1995A&AS..112..525P}, K, SiO \citep{2013MNRAS.434.1469B}, e$^{-}$ \citep{gray2021observation}, H$^{-}$, H, and FeH \citep{2010A&A...523A..58W}. The abundance grid is calculated using the package $\texttt{easyCHEM}$~\citep{2017ApJ...850..150B}. The grid dimensions include $T\in$ [60, 4000]\,K, log\,$P\in[-8,3]$\,bar,  C/O\,$\in[0.1,1.6]$ and [Fe/H]$\in [-2, 3]$, with 100, 100, 20 and 40 equidistant points, respectively. The sampling sizes are thus $\sim$40\,K, $10^{0.11}$\,bar, 0.125\,dex and 0.075 for $T$, $P$, [Fe/H] and C/O, respectively. The C/O ratio varies with the oxygen abundance, once [Fe/H] is set. It is found that the interpolation with a $T$ sampling grid of 100\,K is reasonably accurate to an error budget of a few percent~\citep[e.g., $\sim1.64$\% for water, ][]{Barton2017}, which is less than the estimated uncertainty in the \textit{ab initio} line lists~\citep{Al-Refaie2021}. Given its robustness and rapidness, this approach is widely used in the retrieval analysis of exoplanet atmosphere, although caution should be taken for the atmospheres with one or more of the $T$-$P$-[Fe/H]-C/O values out of the preset boundaries.}

For the retrievals, only the absorption of H$_2$O, CO, CH$_4$, CO$_2$, TiO, VO, FeH are included. Logarithmic uniform prior that vary between $-10$ and 0 are used for each parameter of interest. Collision-induced absorptions (CIAs) of H$_2$-H$_2$ and H$_2$-He \citep{1988ApJ...326..509B, 1989ApJ...336..495B, 2001JQSRT..68..235B, 2002A&A...390..779B}, bound-free and free-free opacity of H$^{-}$, and Rayleigh scattering of H$_2$ and He are also taken into account \citep{1962ApJ...136..690D}. For the FREE model, the mass fractions of each species of investigation, instead of [Fe/H] and C/O, vary independently within the given distributions and boundaries. The same molecules as in the case of equilibrium chemistry are included for retrieval. H$_2$ and He as regarded as filling gases with their mass ratios fixed at the solar values. In addition to the parameters mentioned above, the planet gravity is fixed at $15.8\,{\rm m}\cdot {\rm s}^{-2}$ and radius at 1.646\,$R_{\rm Jup}$ using the literature values from \citet{2015MNRAS.447..711S}.

\ww{Notable offsets in transit depths measured from the ground-based optical data and the space-based NIR data were reported previously in ~\citet[][e.g., ]{Alexoudi2018, Murgas2020, 2021AJ....161....4Y} at the orders of $10^{2-3}$\,ppm. In our case, the measured $J$-band eclipse depth is very close to the HST depth at 1.25$\mu$m\ with only 4\% discrepancy, suggesting that the above-mentioned offset between the ground-based optical depth and space NIR depth seems to be minor (at least if in the same band) in this work. Nevertheless, we still perform vertical offset retrievals to find the most appropriate offset in the range of [$-300,+300$]\,ppm. Comparing the Bayesian evidence ln\,$\mathcal{Z}$ and Bayes factor, we find that the HST data with an offset of $-200$\,ppm could provide the best-fit EQ model, and the HST data an offset of $-50$\,ppm was the best for FREE model. These offsets are smaller or comparable to the measurement uncertainties of the eclipse depths. The yielded abundances for the offset-applied cases are shown in Table~\ref{tab3}, which are well consistent with the un-offsetted ones within $\leqq0.2\sigma$. Therefore, we conclude that applying an offset to the HST NIR data or not does not change the results and conclusions of this work.
}

Overall, both retrievals are well consistent with the data, as presented in Fig.~\ref{Fig:moddel}. The retrieved parameters considered in both equilibrium chemistry and free chemistry models , as well as the $\chi^2_{\nu}$ and Bayesian Evidence for each retrieval are given in Table~\ref{tab3}. The retrieved corner plots are shown in Figs.~\ref{Fig:eq_corner} \& \ref{Fig:free_corner}. Note that $\chi^2_{\nu}$ of the FREE model is slightly smaller than 1, indicating possible overestimates of the data error.

\begin{table}
\centering
\setlength{\tabcolsep}{5pt}

\caption[]{The fixed and retrieved parameters, $\chi^{2}_\nu$ and Bayesian Evidence for each retrieval.\label{tab3}}
 \begin{threeparttable}
 \begin{tabular}{c|c|c|c|c}
 \hline\noalign{\smallskip}
Parameters & EQ$^{a}$ & os=$-200^{b}$ & FREE$^{a}$ & os=$-50^{c}$ \\
 \hline
$R_{\rm p}$($R_{\rm jup}$)       & \multicolumn{4}{c|}{1.646(fixed)} \\
	    \noalign{\smallskip}
log\,g$_{\rm p}$(cgs)  &  \multicolumn{4}{c|}{3.198(fixed)} \\
\noalign{\smallskip}\hline\noalign{\smallskip}
	  
$T_{\rm 1}$(K)      & 
3146$^{+384}_{-135}$& 3384$^{+305}_{-318}$ & 2823$^{+33}_{-30}$& 2788$^{+36}_{-34}$\\
	    \noalign{\smallskip}
$T_{\rm 2}$(K)       & 
2078$^{+224}_{-203}$ &  2178$^{+234}_{-252}$ & 1713$^{+465}_{-734}$& 1250$^{+719}_{-732}$ \\
	    \noalign{\smallskip}
log\,$P_{\rm 1}$(bar)  & 
$-$2.88$^{+0.57}_{-1.52}$  & $-$4.4$^{+1.41}_{-1.09}$ & $-$3.23$^{+0.47}_{-0.97}$ & $-$2.71$^{+0.32}_{-1.22}$\\
	    \noalign{\smallskip}
log\,$P_{\rm 2}$(bar)  & 
1.08$^{+0.46}_{-0.71}$ & 0.87$^{+0.76}_{-0.99}$ & 0.05$^{+0.98}_{-0.98}$ & $-$0.18$^{+1.32}_{-1.15}$\\
	    \noalign{\smallskip}
$[{\rm Fe/H}]$ & 
2.19$^{+0.51}_{-0.63}$ & $2.01^{+0.63}_{-0.89}$ & & \\
	    \noalign{\smallskip}
C/O  & 
1.35$^{+0.14}_{-0.17}$ & $1.28^{+0.21}_{-0.24}$ & 0.86$^{d}$ & \\
	    \noalign{\smallskip}
log$_{10}$(H$^{-}$) 
& & &$-3.72^{+0.29}_{-1.23}$ &
$-3.58^{+0.24}_{-0.64}$\\
	    \noalign{\smallskip}
log$_{10}$(CO$_2$) & & & $-2.66^{+0.92}_{-1.31}$ & $-3.89^{+2.06}_{-3.67}$\\
	    \noalign{\smallskip}
log$_{10}$(H$_2$O) & & & $-6.05^{+2.24}_{-1.96}$ & $-5.94^{+2.40}_{-2.50}$\\
	    \noalign{\smallskip}
log$_{10}$(FeH) & & & $-3.65^{+1.71}_{-3.38}$  & $-4.54^{+2.42}_{-3.40}$\\
	    \noalign{\smallskip}
log$_{10}$(CO) & & & $-4.55^{+2.60}_{-2.99}$ & $-4.80^{+3.15}_{-3.23}$\\
	    \noalign{\smallskip}
log$_{10}$(CH$_4$) & & & $-3.24^{+1.56}_{-3.16}$ & $-4.94^{+2.99}_{-3.13}$\\
	    \noalign{\smallskip}
log$_{10}$(TiO) & & & $-5.93^{+2.20}_{-2.16}$ & $-6.17^{+2.40}_{-2.38}$\\
	    \noalign{\smallskip}
log$_{10}$(VO) & & & $-5.13^{+1.78}_{-2.43}$ & $-5.63^{+2.31}_{-2.65}$\\
	   \noalign{\smallskip}\hline\noalign{\smallskip}

\noalign{\smallskip}\noalign{\smallskip}
$\chi^{2}_\nu$    &1.183 &  1.756 & 0.9199 & 1.348  \\
\noalign{\smallskip}
ln\,$\mathcal{Z}$ & $971.9\pm0.4$ & $972.5\pm0.0$ & $970.1\pm1.0$ & $972.7\pm0.5$\\
\hline\noalign{\smallskip}

\end{tabular}
\begin{tablenotes}
\item[a] The degrees of freedom (D.o.F) for the EQ and FREE retrievals are 6 and 12, respectively. 
\ww{
\item[b] The EQ retrieval after applying an offset of $-200$\,ppm to the HST data. 
\item[c] The FREE retrieval after applying an offset of $-50$\,ppm to the HST data. 
\item[d] This is a rough estimate.
\\
}
\end{tablenotes}
\end{threeparttable}
\end{table}

\subsection{Results and discussion}
\label{sec:discussion}
\begin{figure*}

\centering     
\subfigure[]{\label{Fig9:a}\includegraphics[width=85mm]{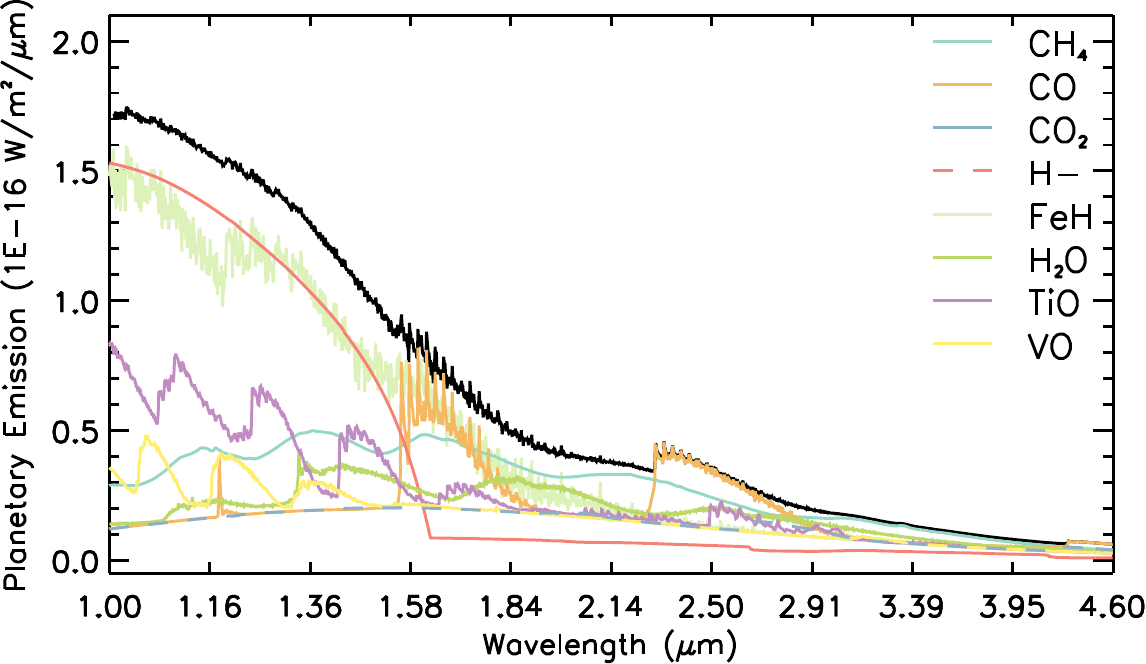}}
\subfigure[]{\label{Fig9:b}\includegraphics[width=85mm]{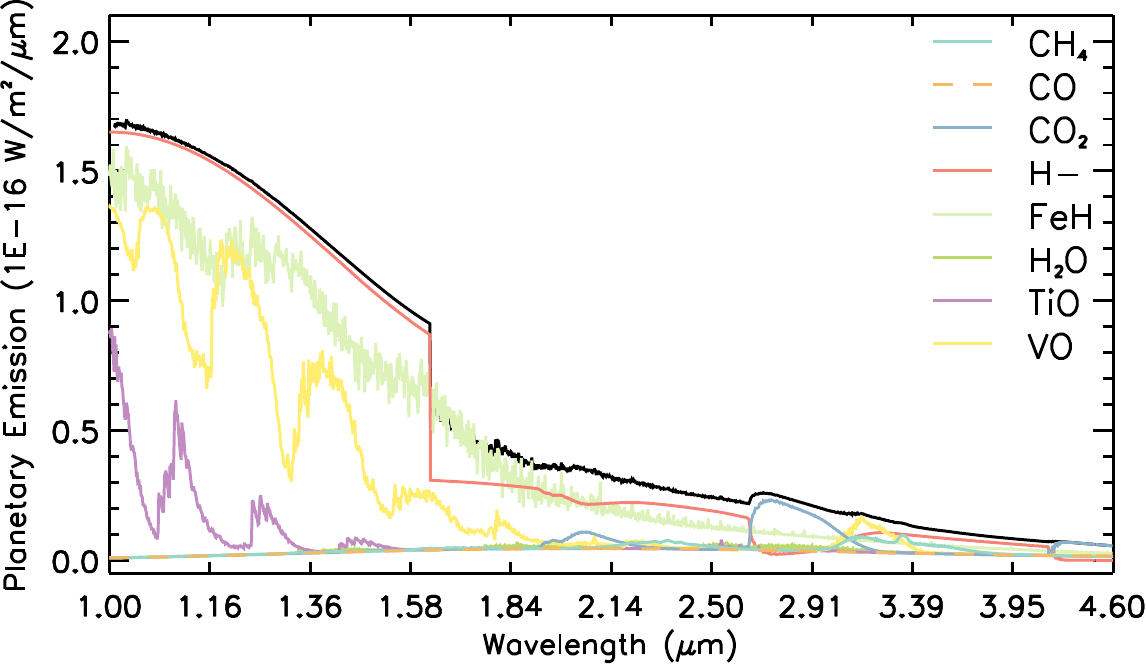}}
\caption{The emission spectra of the most dominant species in our EQ and FREE models. The left panel shows the EQ model, and the right panel shows the FREE model. }

\label{Fig:molecular}
\end{figure*}

The best-fit EQ model is shown in Fig.~\ref{Fig:moddel} as the yellow line with $\chi^{2}_\nu=1.183$ comparing in general well with the observation. Nevertheless, the several reddest HST data points and our $K_{\rm s}$ data stand apart from the model prediction by $\gtrsim 1\sigma$. As listed in Table~\ref{tab3}, the best-fit [Fe/H] and C/O ratio are $2.19^{+0.51}_{-0.63}$\,dex and $1.35^{+0.14}_{-0.17}$, respectively, indicating a metal-rich and carbon-rich atmosphere. The significantly high metallicity and C/O ratio in the atmosphere of WASP103\,b result in large accumulation of FeH and CO. Given that FeH has strong blue-slope features in the $J$ band regime, and CO has a strong blue-slope emission band from $\sim 2.26-3\,\micron$~(cf. Fig.~\ref{Fig:molecular}), their combined emissivity contribution compensates a bit for the relative low emission at 1.62\,\micron\ and the $K_{\rm s}$ band, although still not good enough. This implies that the equilibrium chemistry assumption may not apply to the dayside atmosphere of WASP103\,b, as found for many other planets 
\ww{\citep{mcgruder2022access, mansfield2022confirmation, 2021A&A...656A..90K, 2021AJ....162...37R}}. 

Indeed, as shown in Fig.~\ref{Fig:moddel} as the blue line, the FREE model spectrum fits the observation better than the EQ model does, with $\chi^{2}_\nu=0.92$. The best-fit mass fractions are listed in the third column of Table~\ref{tab3}. The most abundant molecules are CO$_2$, CH$_4$, FeH and H$^{-}$, which have mass fractions larger than 10$^{-4}$. Taking into account them and CO, we find a C/O number ratio of 0.86, higher than the solar value, although not as extreme as the EQ model result. Looking into the Fig.~\ref{Fig:molecular}(a), the overabundance of H$^{-}$ and the discontinuity at $\sim1.6\,\micron$ seem to be the main reason to account for the sudden decrease of the reddest HST data and our $K_{\rm s}$ data. It is usually the case for classic HJs that H$_2$O, H$^{-}$, and hydrides/oxides (TiO, VO and FeH) are the major species that contribute to the opacity over the HST/WFC3 wavelength range. For WASP103\,b, however, the dayside temperature ($T>1710$\,K) is so high that H$_2$O is dissociated, and H atoms in the higher altitude bond with free electrons and become H$^{-}$, which devotes opacity to the atmosphere bringing a continuum absorption\citep{1988A&A...193..189J} and drags down suddenly the spectrum at $\sim 1.6$\,\micron, as presented in Fig.~\ref{Fig:moddel}. CO$_2$ emission is visible near the wavelength of 2.3\,\micron and 4.5\,\micron, yielding a mixing ratio of CO$_2$ of $-2.66^{+0.92}_{-1.31}$.

In order to verify the reliability of \ww{the detected molecules}, we follow the approach introduced in \citet{2013ApJ...778..153B, 2022AJ....164..124K}. We remove one molecule at a time and redo the retrieval, and calculate for each model the Bayesian evidence $\mathcal{Z}$, which are listed in Table~\ref{tab4}. By comparing $\mathcal{Z}$ between the full model and the model without a given species, we could investigate whether the observation favours the model including that species. The Bayes factor\,\citep{doi:10.1080/01621459.1995.10476572} of CO$_2$, H$^{-}$, \ww{CH$_4$, FeH and TiO} are large enough ($>6$), indicating that these molecules are largely favored by the data to be included in the model. As CO doesn't have strong features in the observed wavelength coverage, the retrieval analysis with CO excluded is similar to the full model. 

It is worthy to mention that for each model we investigate in this work, EQ or FREE, or with a certain molecule removed, the retrieved T-P profile possess a thermal inversion layer. This is in hand with the findings in~\citet{cartier2016near, 2018AJ....156...17K, changeat2022spectroscopic}. For the atmospheric composition, \citet{cartier2016near} claimed a solar metallicity with C/O$>$1, while \citet{2018AJ....156...17K} derived a moderately enhanced metallicity ($23_{-13}^{+29}\times$ solar) and a C/O ratio of 0.9, consistent with our results. Our EQ model yields a [Fe/H]=$2.19^{+0.51}_{-0.63}$\,dex and C/O ratio of $1.35^{+0.14}_{-0.17}$. The FREE model does not set [Fe/H] or C/O as retrieved parameters, and thus can not place direct constraint on these two parameters. However, we estimate a C/O number ratio of $\sim0.96$ from the mass abundance of the most dominate species, which is close to the value obtained by \citet{2018AJ....156...17K}. Interestingly, \citet{changeat2022spectroscopic} used the same HST and Spitzer data, and yielded a solar metallicity, and a supersolar C/O ratio. Therefore, although whether the atmosphere of WASP-103\,b is metal-rich is controversial, it is all agreed that the atmosphere has a super-solar C/O ratio. It is also notable that both our work and \citet{changeat2022spectroscopic} detected a large accumulation of H$^-$ and FeH on the dayside of WASP-103\,b.

\begin{table}
\centering
\setlength{\tabcolsep}{5pt}

\caption[]{The Bayesian evidence difference of the model with a certain molecule not included as compared to the full model for the FREE retrievals.\label{tab4}}

 \begin{tabular}{c|c|c}
 \hline\noalign{\smallskip}
Model & $\Delta \rm In(\mathcal{Z})$ & Bayes factor\\
 \hline
no CO$_2$   & $+4.5\pm$1.02& 90.02\\
no H$^{-}$  & $+3.7\pm$1.28& 40.45\\
no CH$_4$   & $+2.8\pm$1.22& 16.5\\
\ww{no TiO} & $+2.1\pm$1.11& 8.17\\
\ww{no FeH}      & $+1.9\pm$1.22& 6.69\\
no H2O      & $+1.0\pm$1.22& 2.72\\
\ww{no VO}       & $+0.9\pm$1.34& 2.45\\
no CO       & $-2.4\pm$1.12& 0.09\\
\hline\noalign{\smallskip}

\end{tabular}

\end{table}

\section{Summary}
\label{sec:summary}

We took two secondary eclipses observations of the hot Jupiter WASP-103\,b with CFHT/WIRCam in $J$ and $K_{\rm s}$, and reanalyzed the data observed by \citet{delrez2018high} in $K_{\rm s}$ band. After correcting the contamination from a faint nearby star, the derived eclipse depths are 0.131$\pm$0.022\% and 0.135$\pm$0.027\% in $J$ and $K_{\rm s}$, respectively. Our $J$ band eclipse depth is consistent with \citet{cartier2016near, 2018AJ....156...17K} within 1$\,\sigma$. The HST/WFC3 emission spectra covering 1.15–1.65\,\micron\ and Spitzer/IRAC broadband photometry (3.6 and 4.6\,\micron\ bands) are combined with our $J$ \& $K_{\rm s}$ measurements to study the atmopshere of WASP103\,b.

The retrieval analysis performed in this study \ww{assuming equilibrium or free chemistry yield in general similar conclusions that the atmosphere is quite metal rich and carbon rich with respect to its host star and the sun. However, the calculated C/O ratios are quite different, which should be due to two facts. One is that the retrieved T-P profiles are different. The other is that the fractions of each contributing molecules are different, as could be observed in Fig.~\ref{Fig:molecular}. For example, the carbon-bearing molecules CO and CO$_2$ in the EQ model are much more prominent than those in the FREE model, while the situation for TiO and VO are on the contrary. This two changes naturally lead to the obviously different C/O ratios between the EQ and FREE models. We note that assuming equilibrium chemistry might not be suitable, as many UHJ appear not at the chemical equilibrium state due to the observed enrichment in refractory elements. On the other side, the use of FREE model could also be risky for some cases as the fractions of individual species could be nonphysical.} The dayside atmosphere is best fit with a thermal inversion on the dayside, which is likely associated with the strong optical absorber FeH, instead of TiO or VO, given the high temperature in this planet. A large abundance of FeH and H$^{-}$ is favoured by the free chemistry retrieval, in agreement with the discovery in \cite{changeat2022spectroscopic}. A supersolar C/O ratio (0.86 or 1.5) is recommended by our retrieval analyisis, which is consistent with the conclusions from \citet{cartier2016near, 2018AJ....156...17K, changeat2022spectroscopic}. Our analysis suggests that the atmosphere has a supersolar metallicity of $\sim$2.19\,dex, confirming the previous finding by \citet{2018AJ....156...17K}, although \citet{cartier2016near, changeat2022spectroscopic} obtained a solar metallicity. We emphasize here that adding a filter at $\sim$2.2\,\micron\ to the HST/WFC3 data may provide crucial information on the chemical compositions, particularly carbon oxides CO$_2$ and CO. Future emission spectroscopy observations of WASP103\,b with JWST, for example, are needed to yield more reliable metallicity and C/O ratio measurements, as well as mass fractions of individual atoms/molecules in the atmosphere.

\section*{Acknowledgements}

This research is supported by the National Natural Science Foundation of China grants No. 11988101, 42075123, 42005098, 62127901, the National Key R\&D Program of China No.~2019YFA0405102, the Strategic Priority Research Program of Chinese Academy of Sciences, Grant No.~XDA15072113, the China Manned Space Project with NO. CMS-CSST-2021-B12. Y.-Q.S, M.Z., J.-S.H. are supported by the Chinese Academy of Sciences (CAS), through a grant to the CAS South America Center for Astronomy (CASSACA) in Santiago, Chile. The authors thanks the referee for the constructive and useful comments and suggestion. 

This research uses data obtained through the Telescope Access Program (TAP), which has been funded by the National Astronomical Observatories of China, the Chinese Academy of Sciences (the Strategic Priority Research Program “The Emergence of Cosmological Structures” Grant No. XDB09000000), and the Special Fund for Astronomy from the from the Ministry of Finance. Two observational nights (2015 May 28 and June 9) with the WIRCam on the 3.6 m CFHT telescope were distributed to us for scientific studies of exoplanetary atmosphere via the TAP. 

This paper makes use of EXOFAST (Eastman et al. 2013) as provided by the NASA Exoplanet Archive, which is operated by the California Institute of Technology, under contract with the National Aeronautics and Space Administration under the Exoplanet Exploration Program

\section*{Data Availability}


The CFHT data is accessible via the Canadian Astronomy Data Centre portal at
\href{https://www.cadc-ccda.hia-iha.nrc-cnrc.gc.ca/en/cfht/}{https://www.cadc-ccda.hia-iha.nrc-cnrc.gc.ca/en/cfht/}. 
The {\tt\string EXOFAST} code is accessible following the instruction via \href{https://astroutils.astronomy.osu.edu/exofast/pro/exofast/README}{https://astroutils.astronomy.osu.edu/exofast/pro/exofast/README}. 
The {\tt\string petitRADTRANS} retrieval modelling code and associated python scripts for parameter analysis and plotting are available via \href{https://petitradtrans.readthedocs.io/en/latest/content/installation.html}{https://petitradtrans.readthedocs.io/en/latest/content/installation.html}.



\bibliographystyle{mnras}
\bibliography{reference_103b} 




\appendix

\section{Additional figures}

\begin{figure*}
\subfigure[]{\includegraphics[width=55mm]{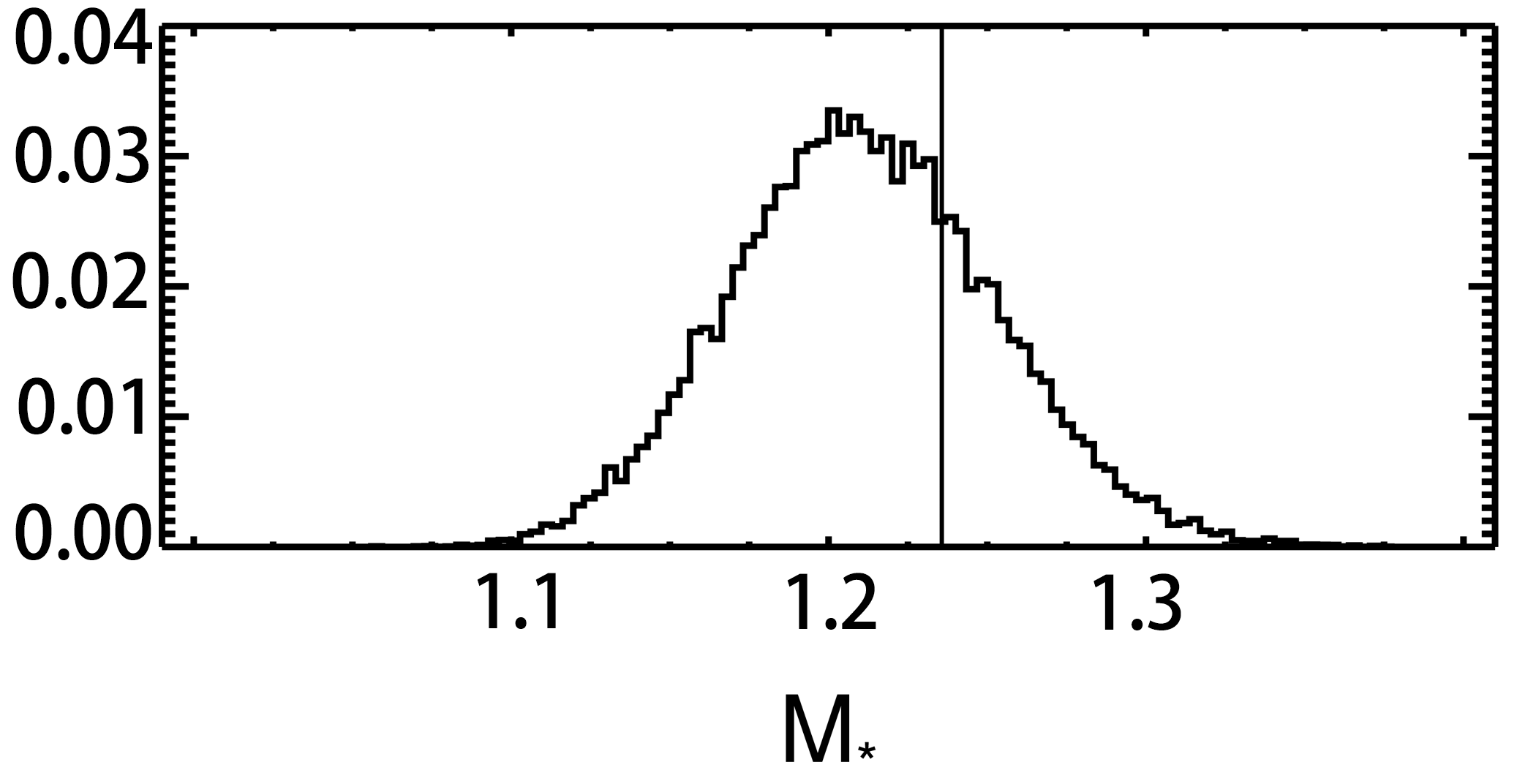}}
\subfigure[]{\includegraphics[width=55mm]{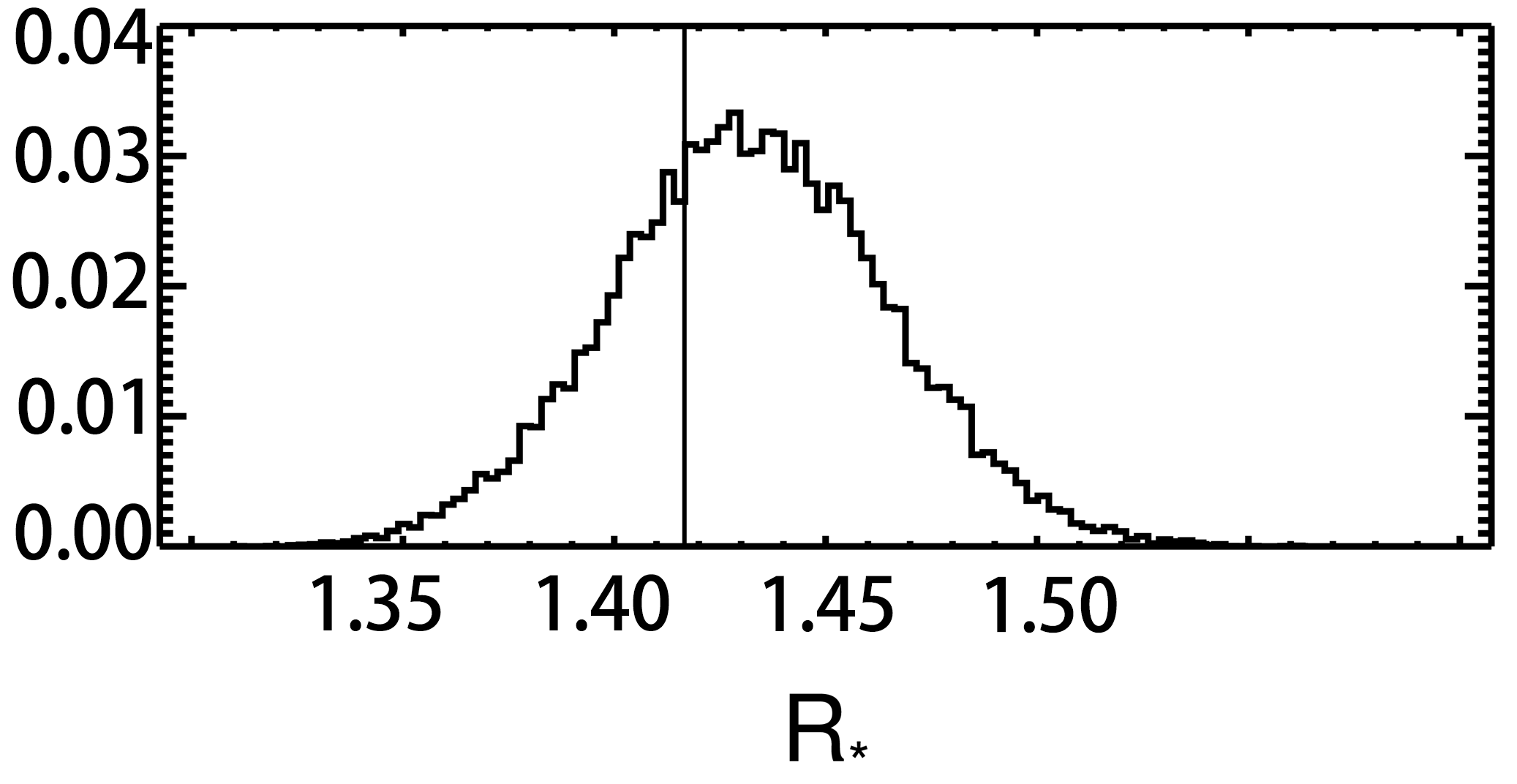}}
\subfigure[]{\includegraphics[width=55mm]{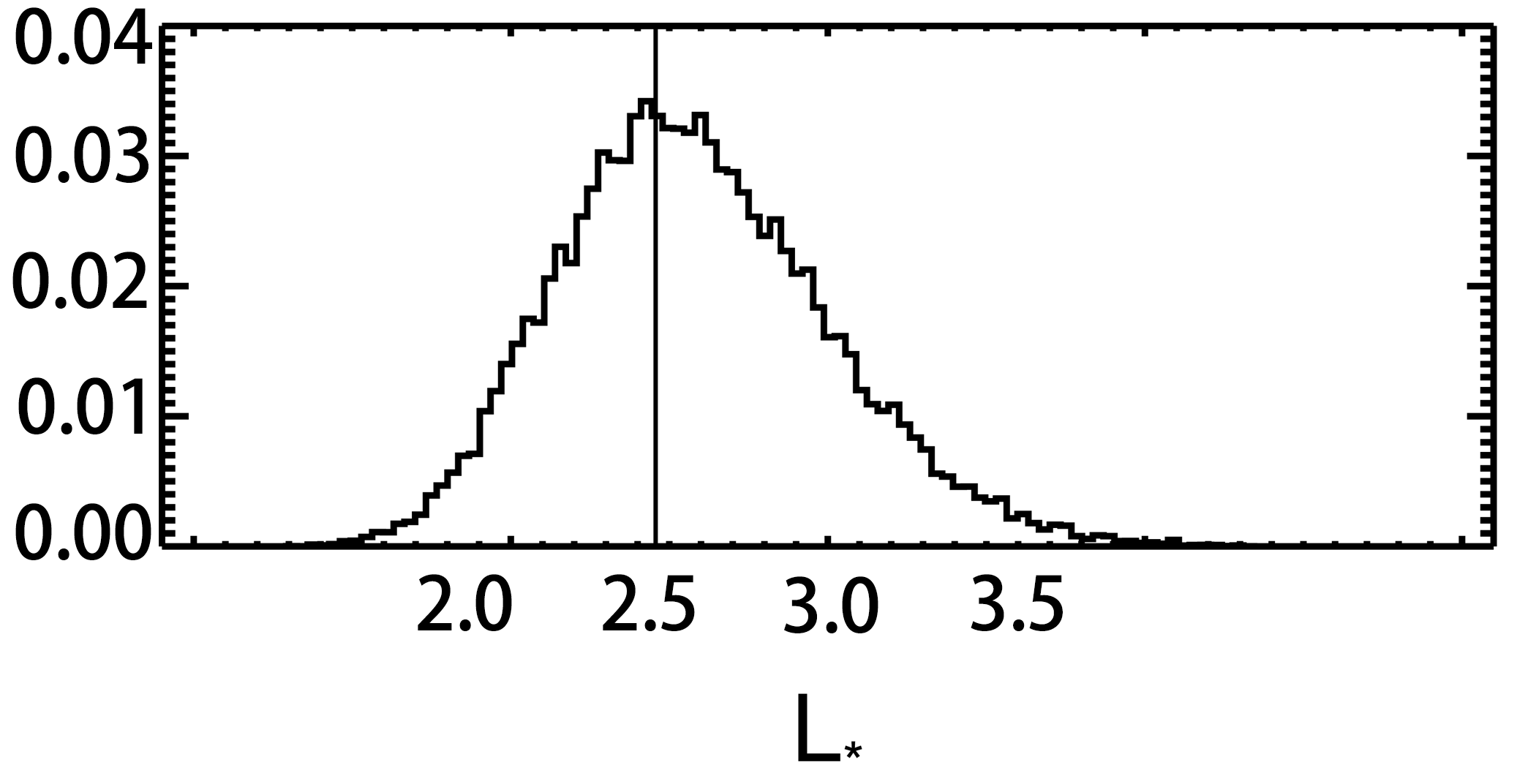}}
\subfigure[]{\includegraphics[width=55mm]{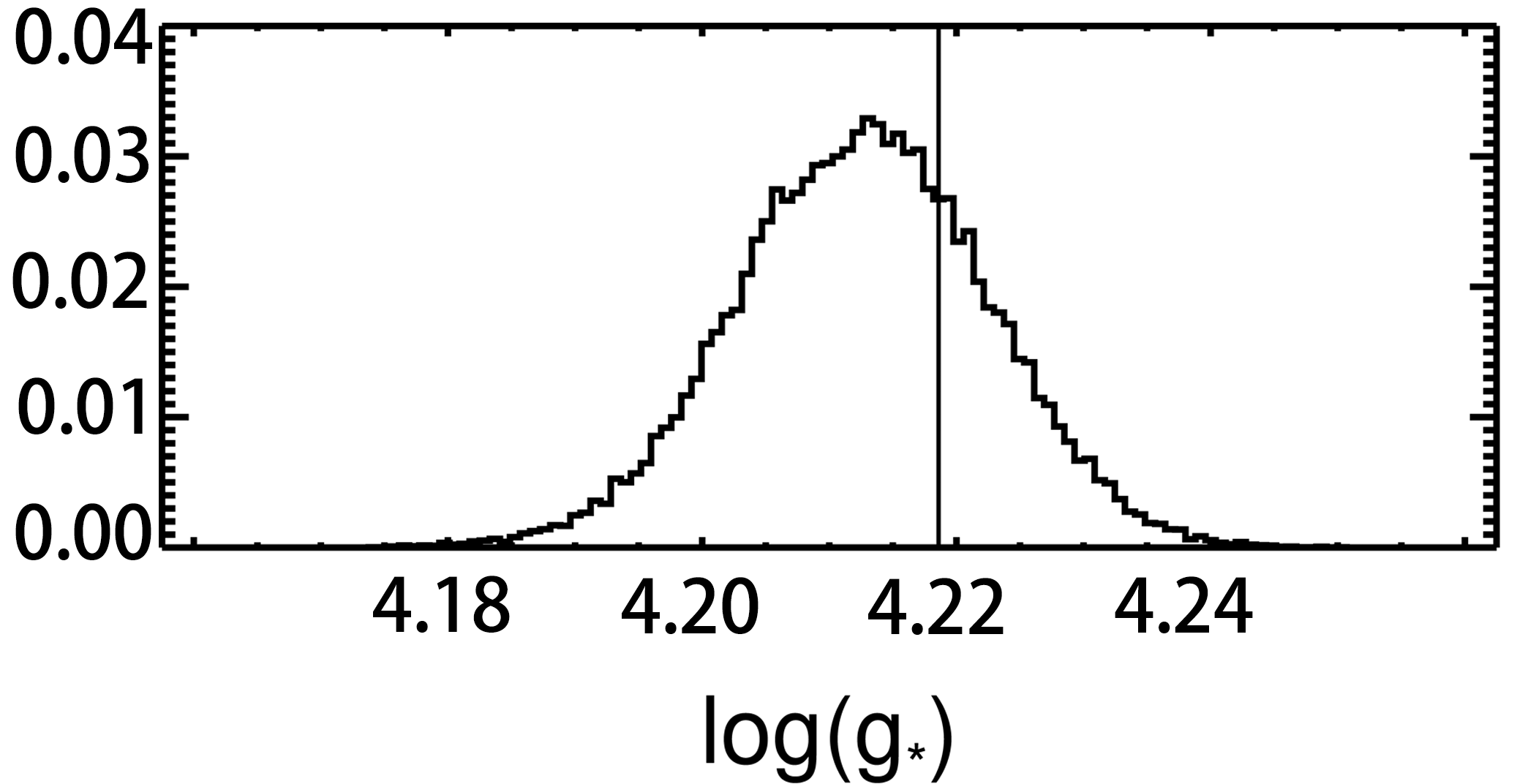}}
\subfigure[]{\includegraphics[width=55mm]{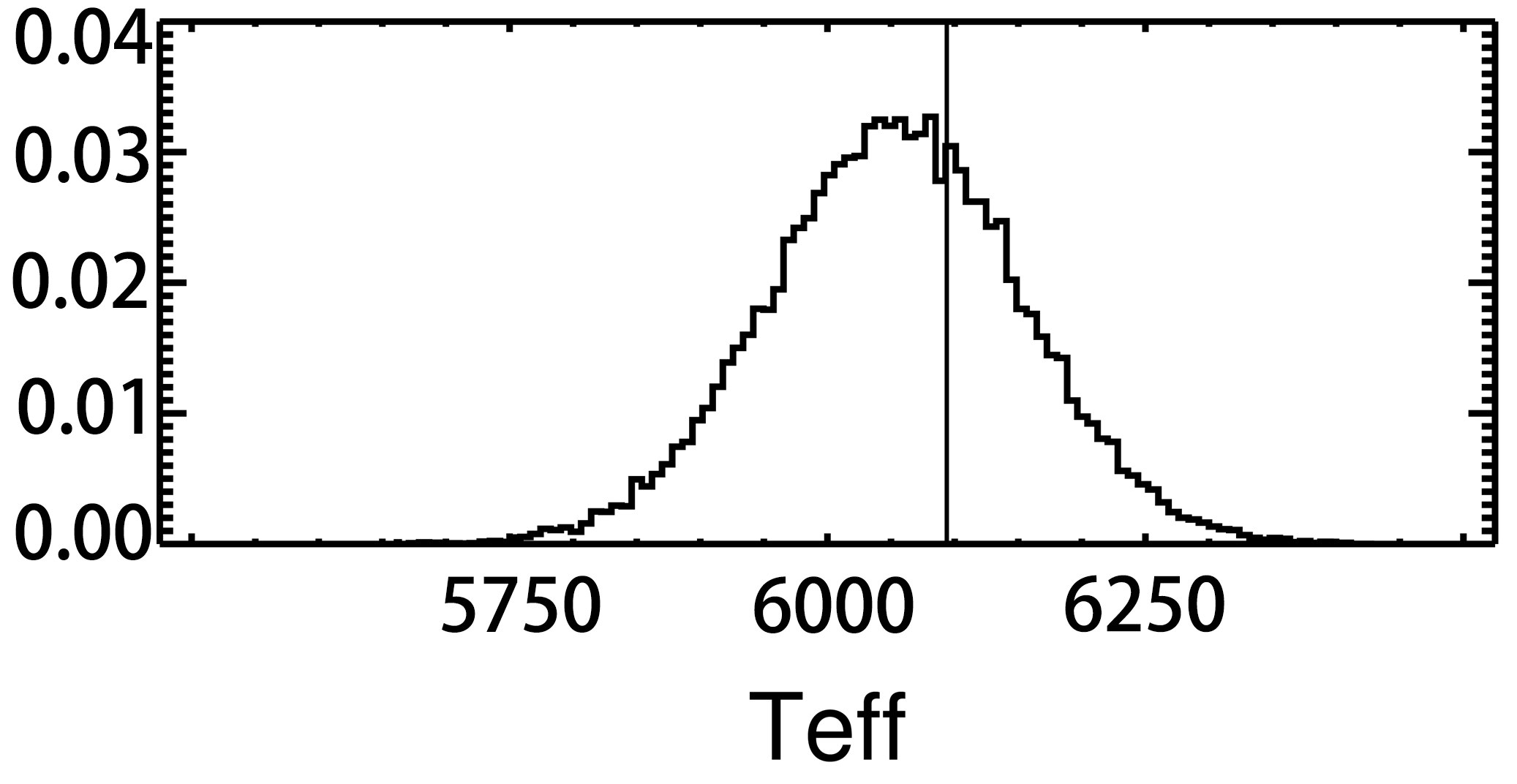}}
\subfigure[]{\includegraphics[width=55mm]{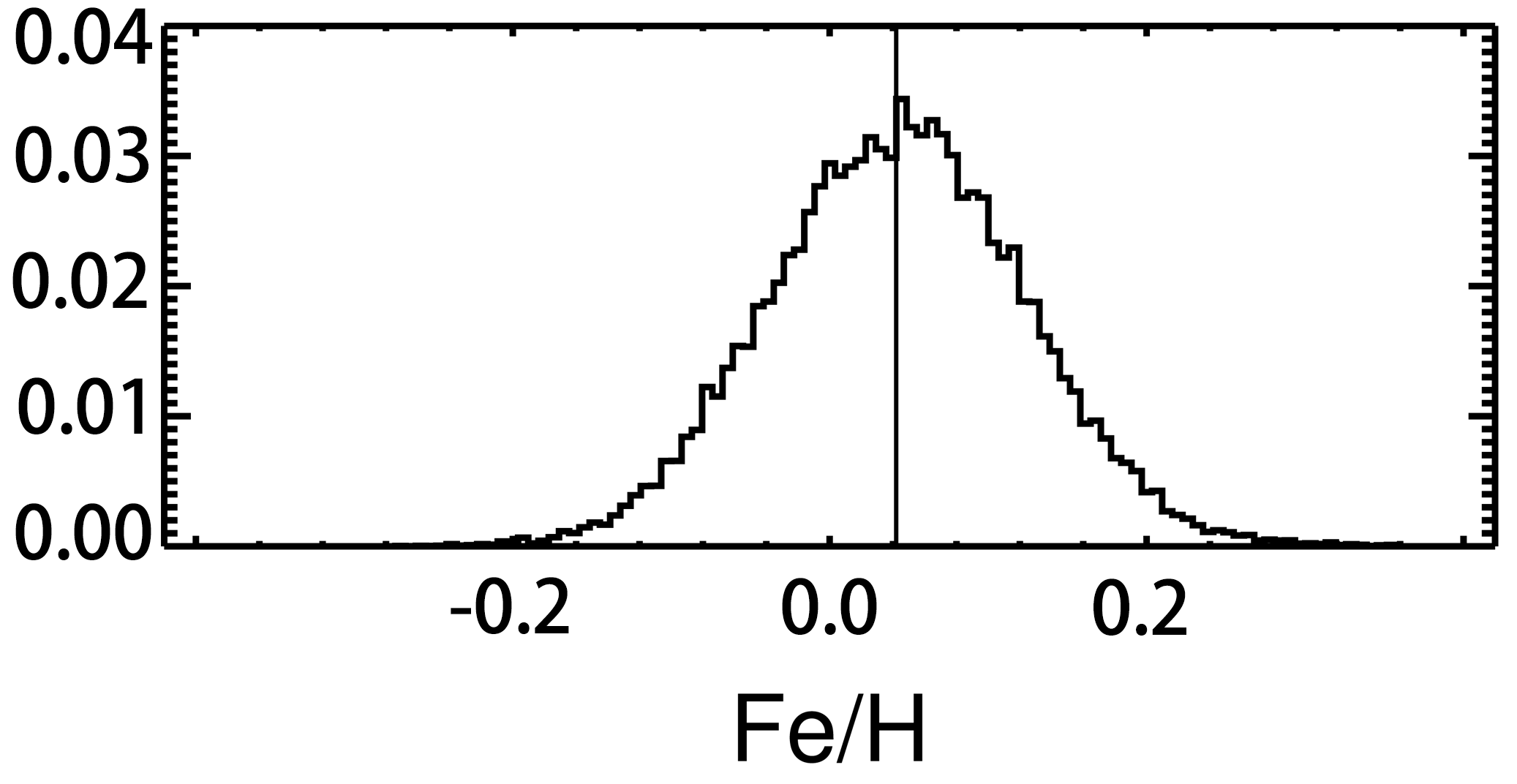}}
\subfigure[]{\includegraphics[width=55mm]{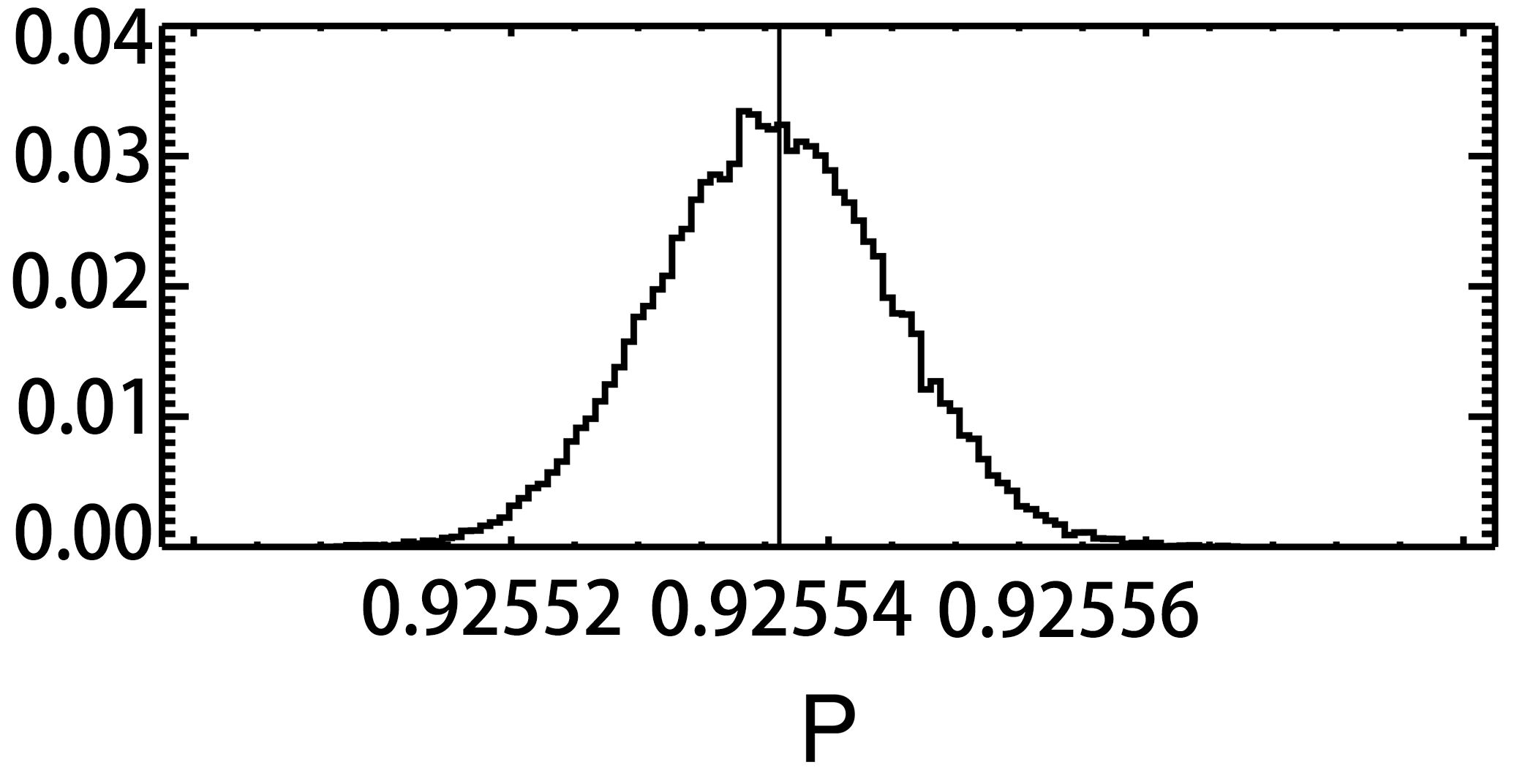}}
\subfigure[]{\includegraphics[width=55mm]{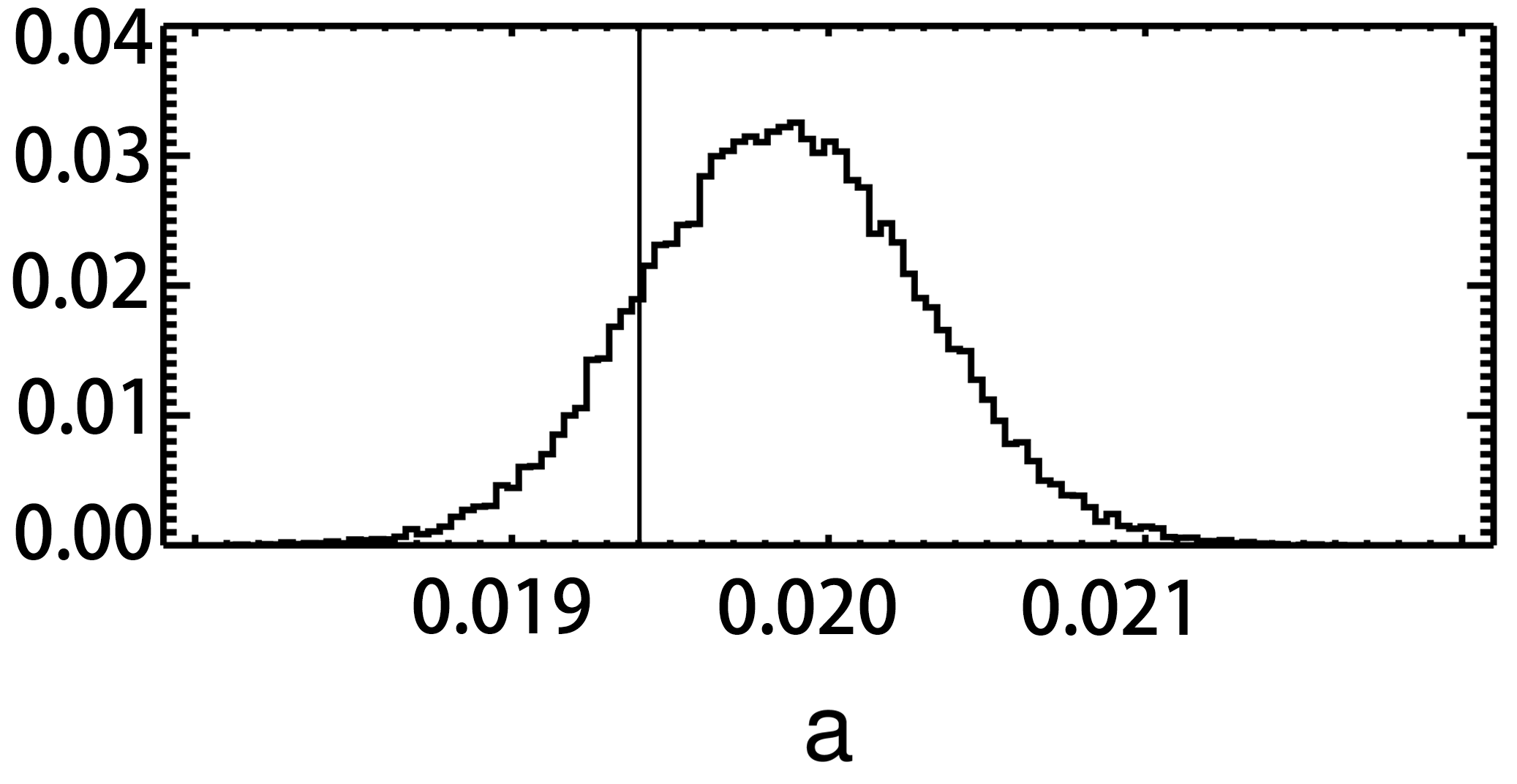}}
\subfigure[]{\includegraphics[width=55mm]{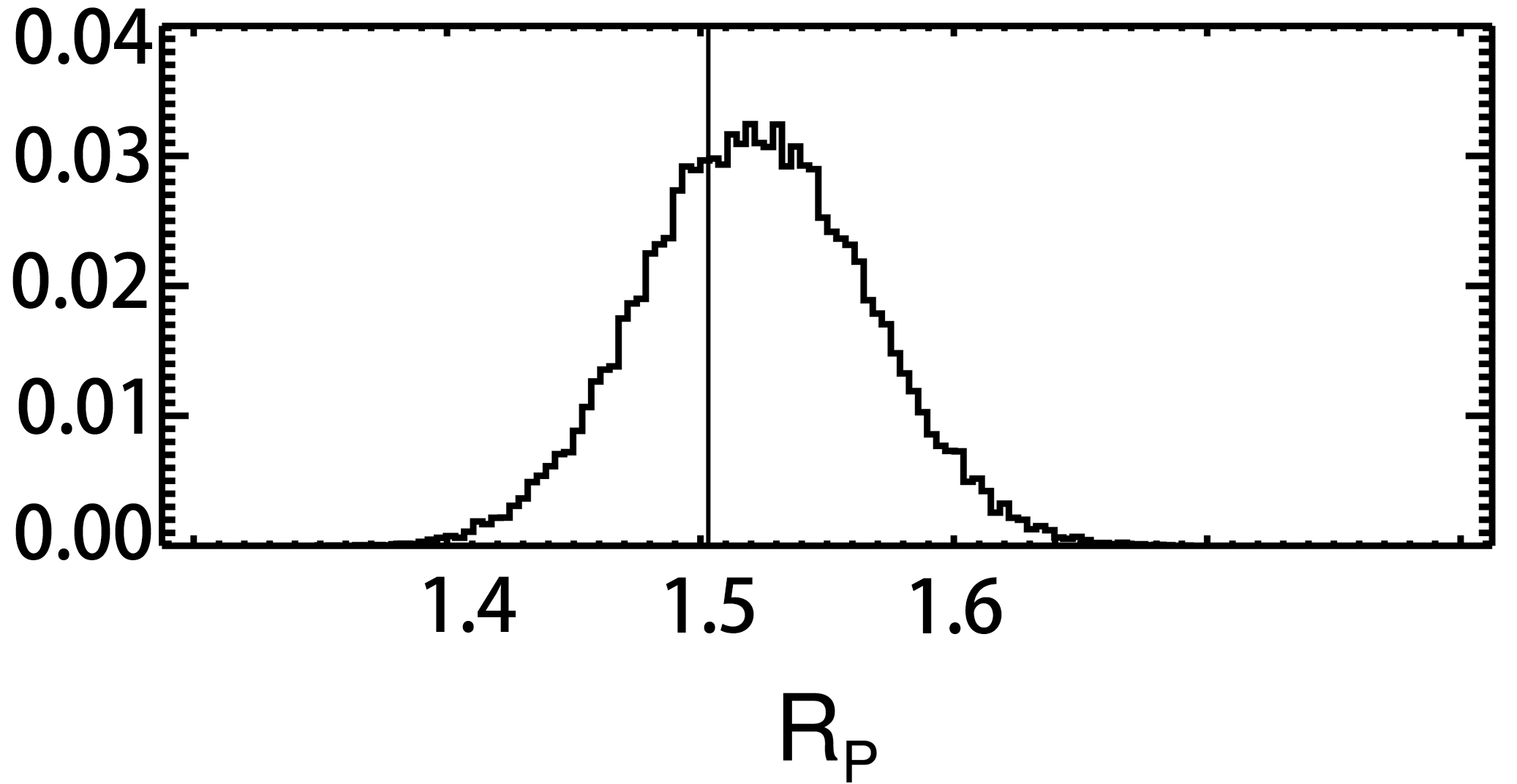}}
\subfigure[]{\includegraphics[width=55mm]{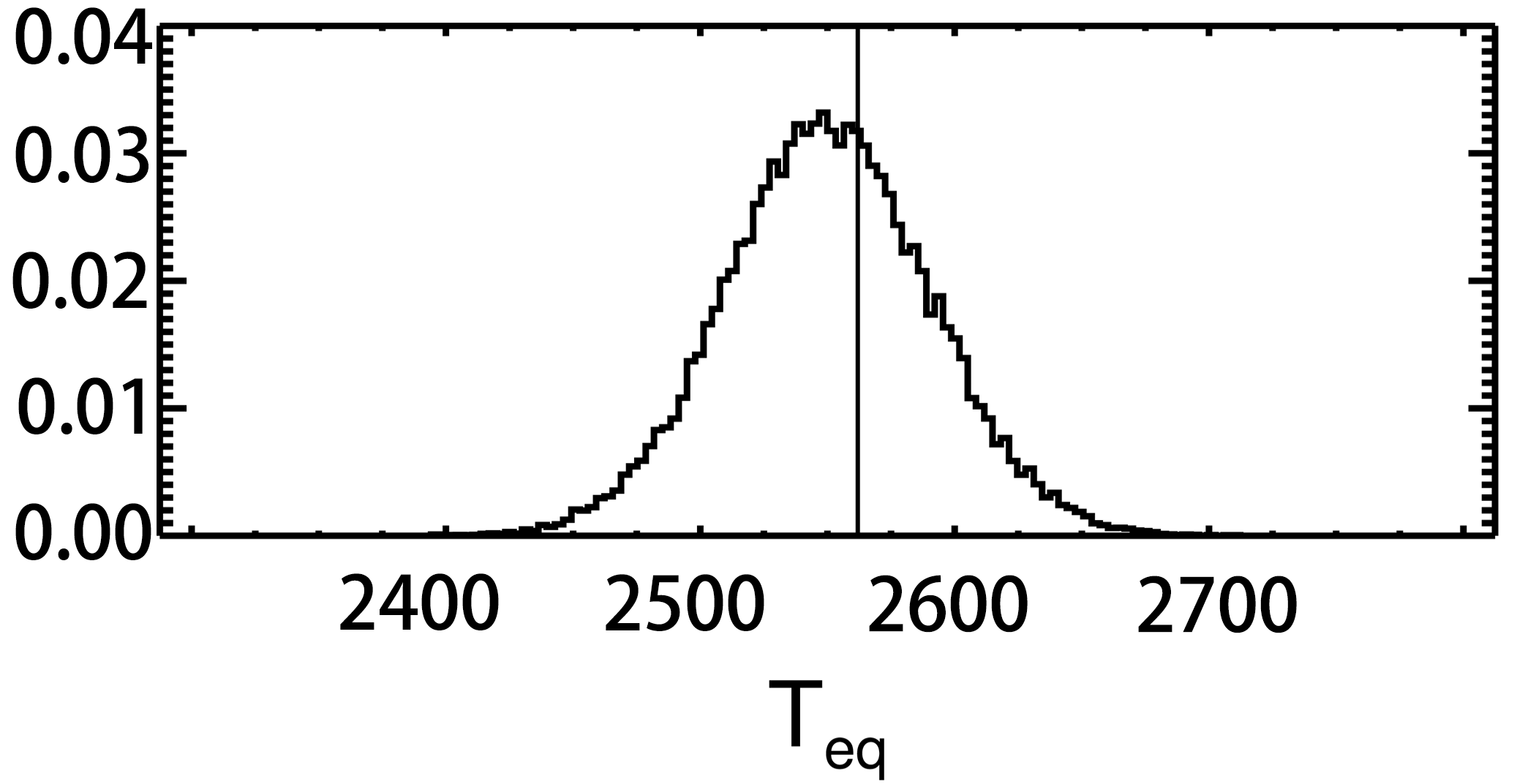}}
\subfigure[]{\includegraphics[width=55mm]{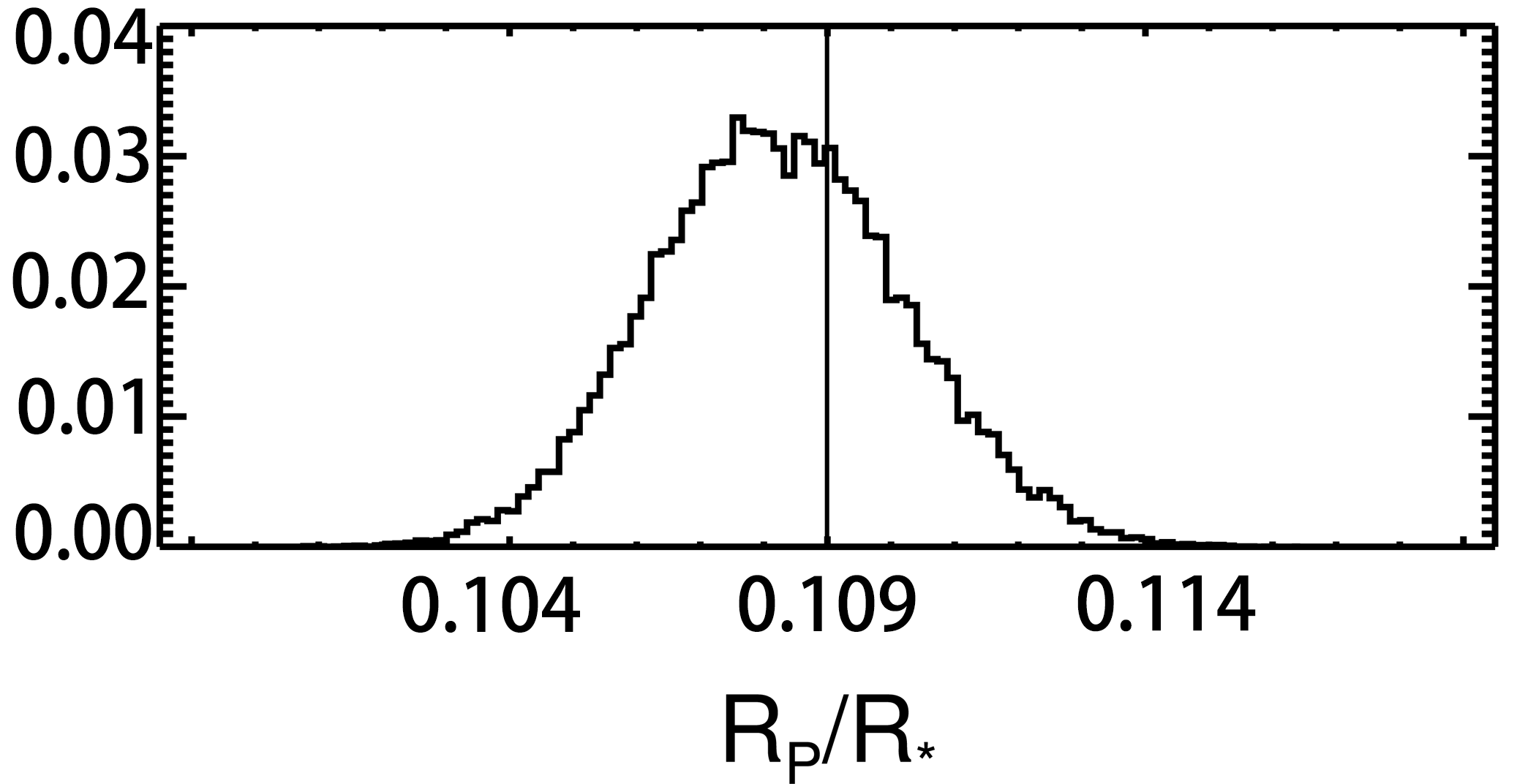}}
\subfigure[]{\includegraphics[width=55mm]{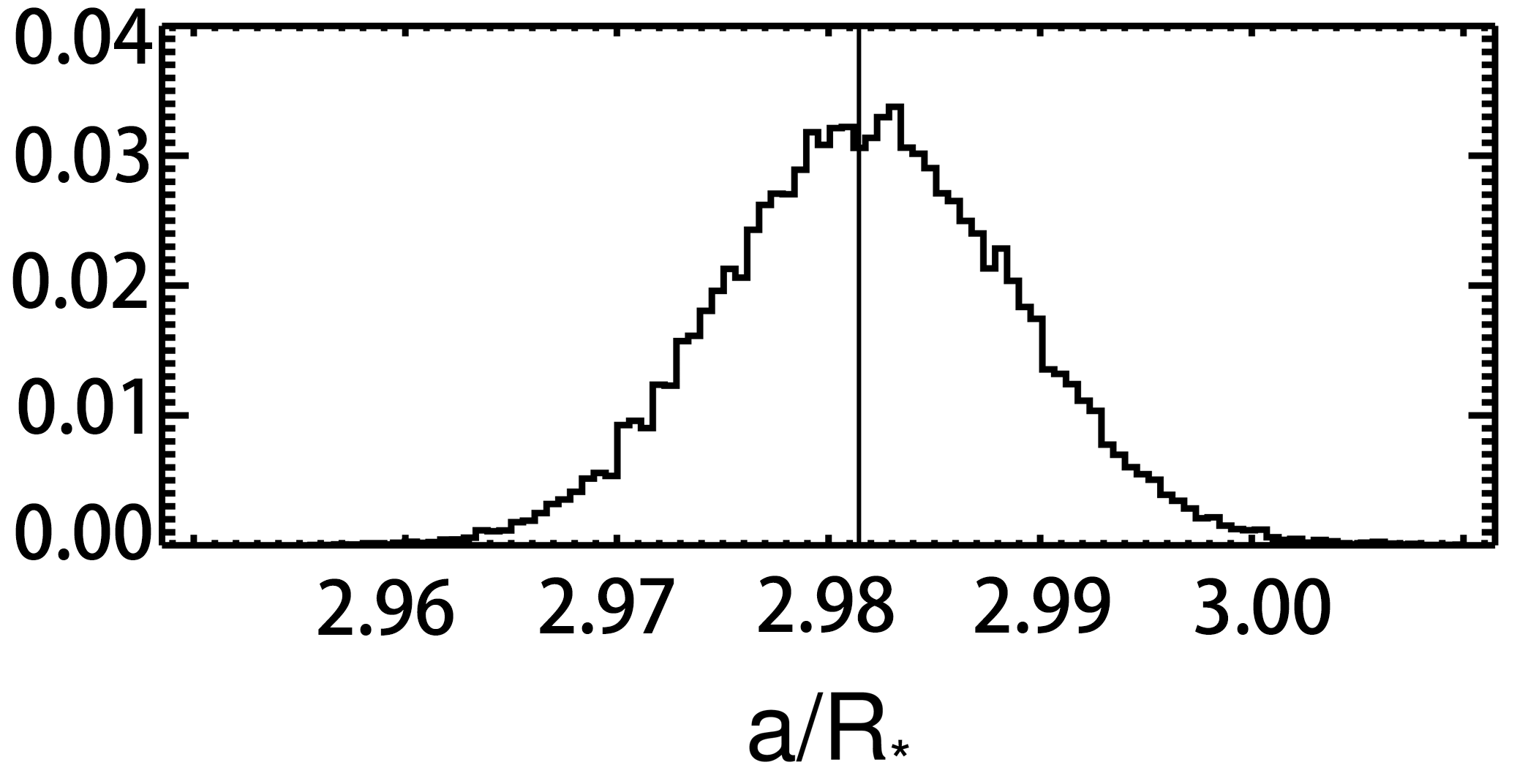}}
\subfigure[]{\includegraphics[width=55mm]{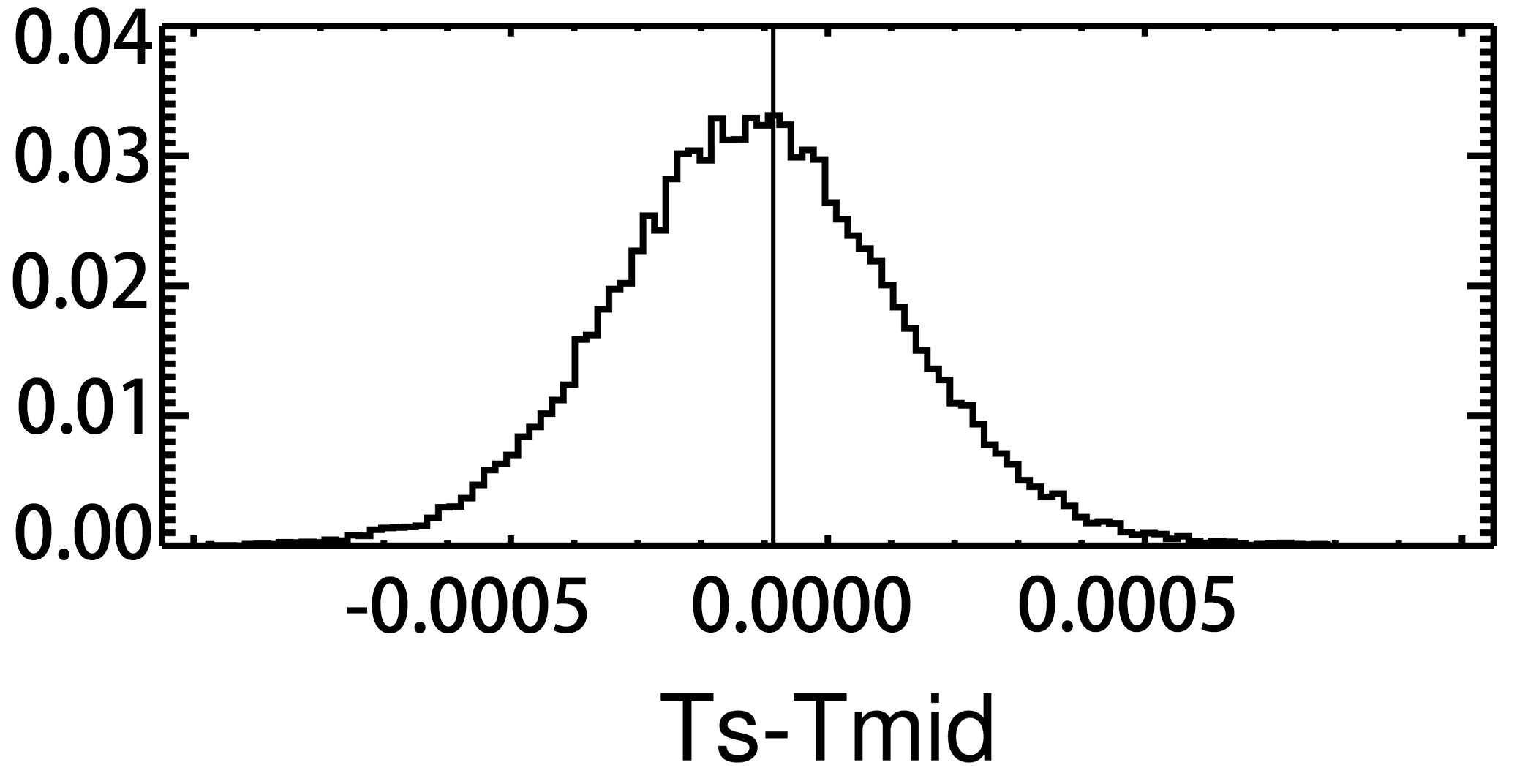}}
\subfigure[]{\includegraphics[width=55mm]{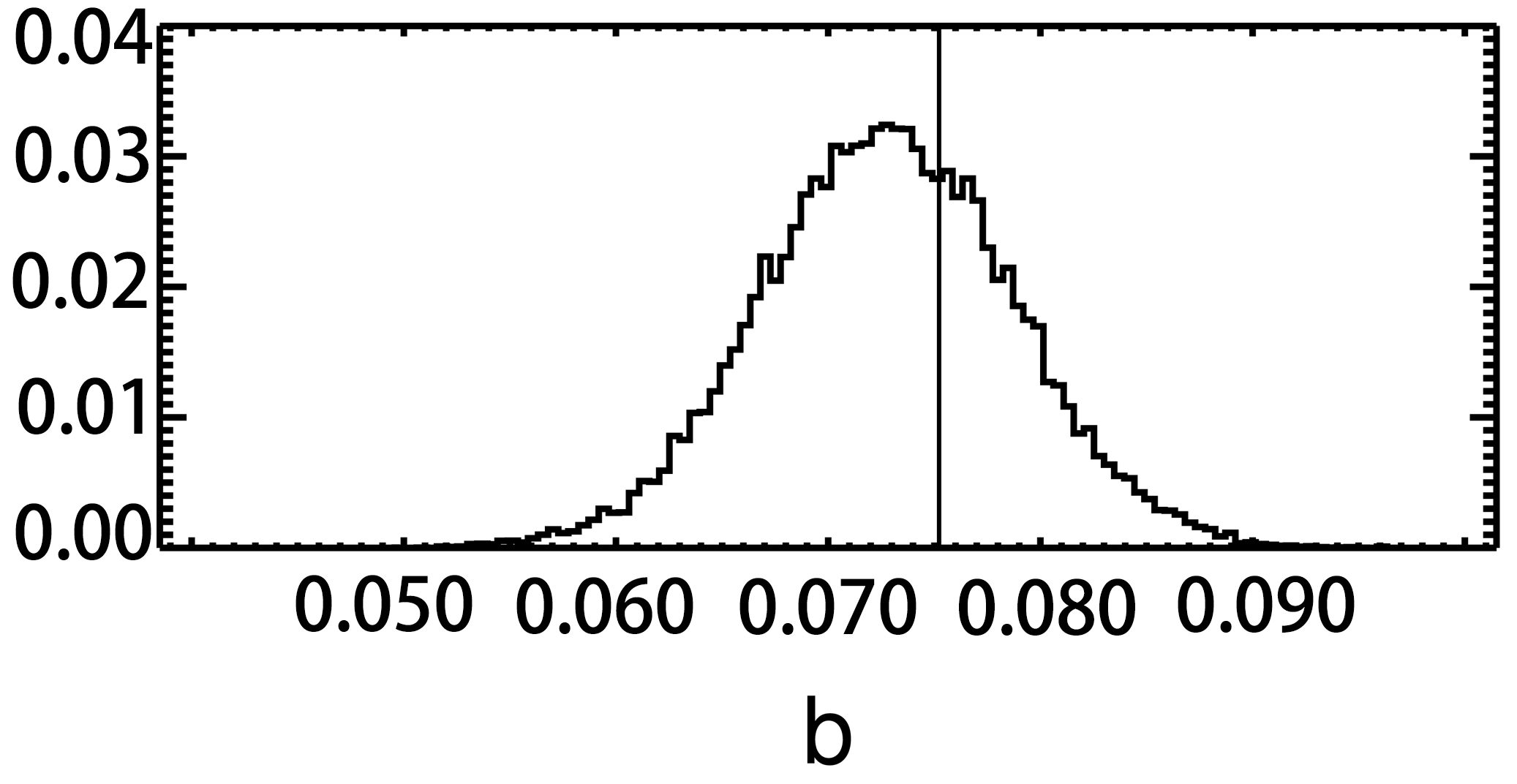}}
\subfigure[]{\includegraphics[width=55mm]{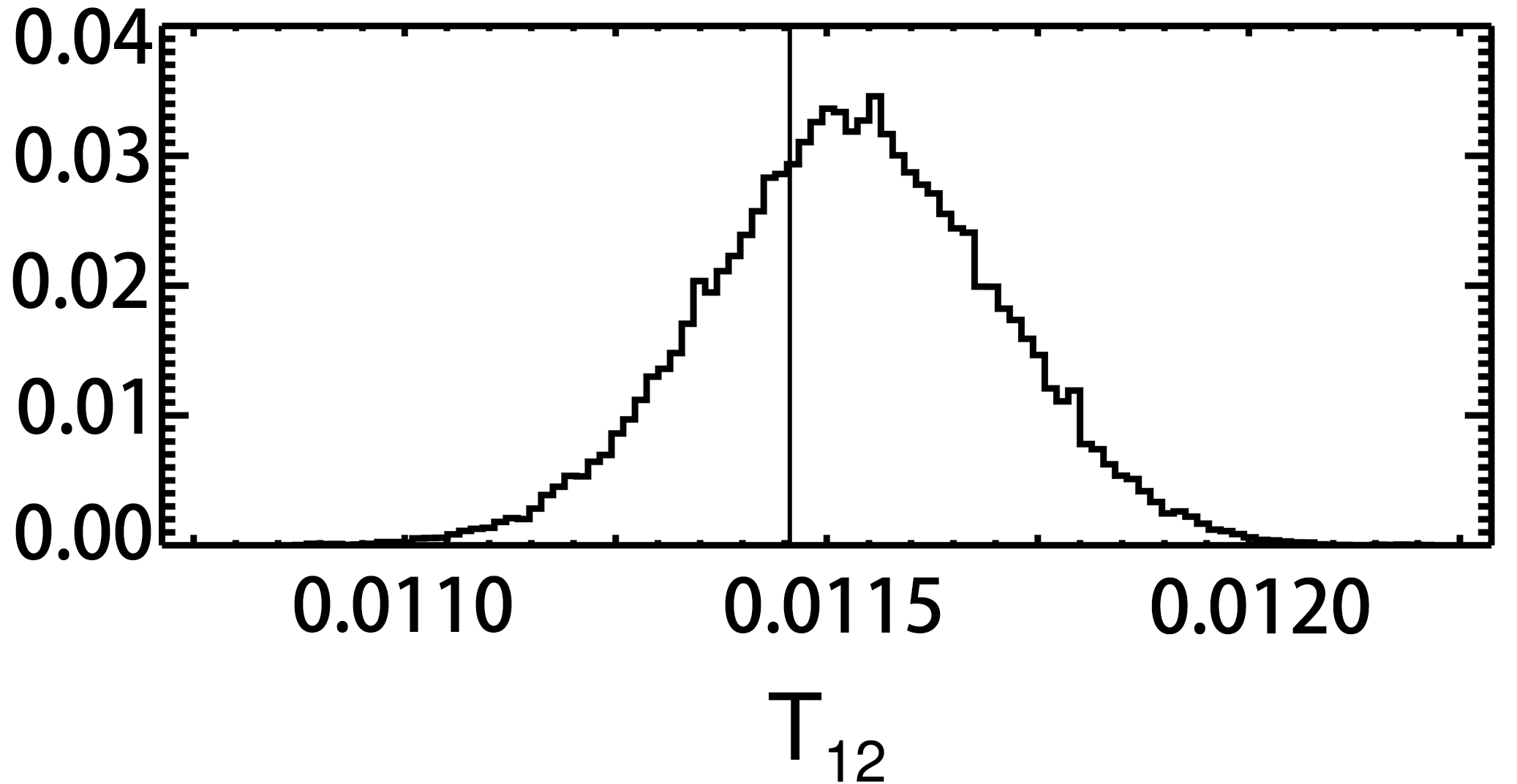}}
\subfigure[]{\includegraphics[width=55mm]{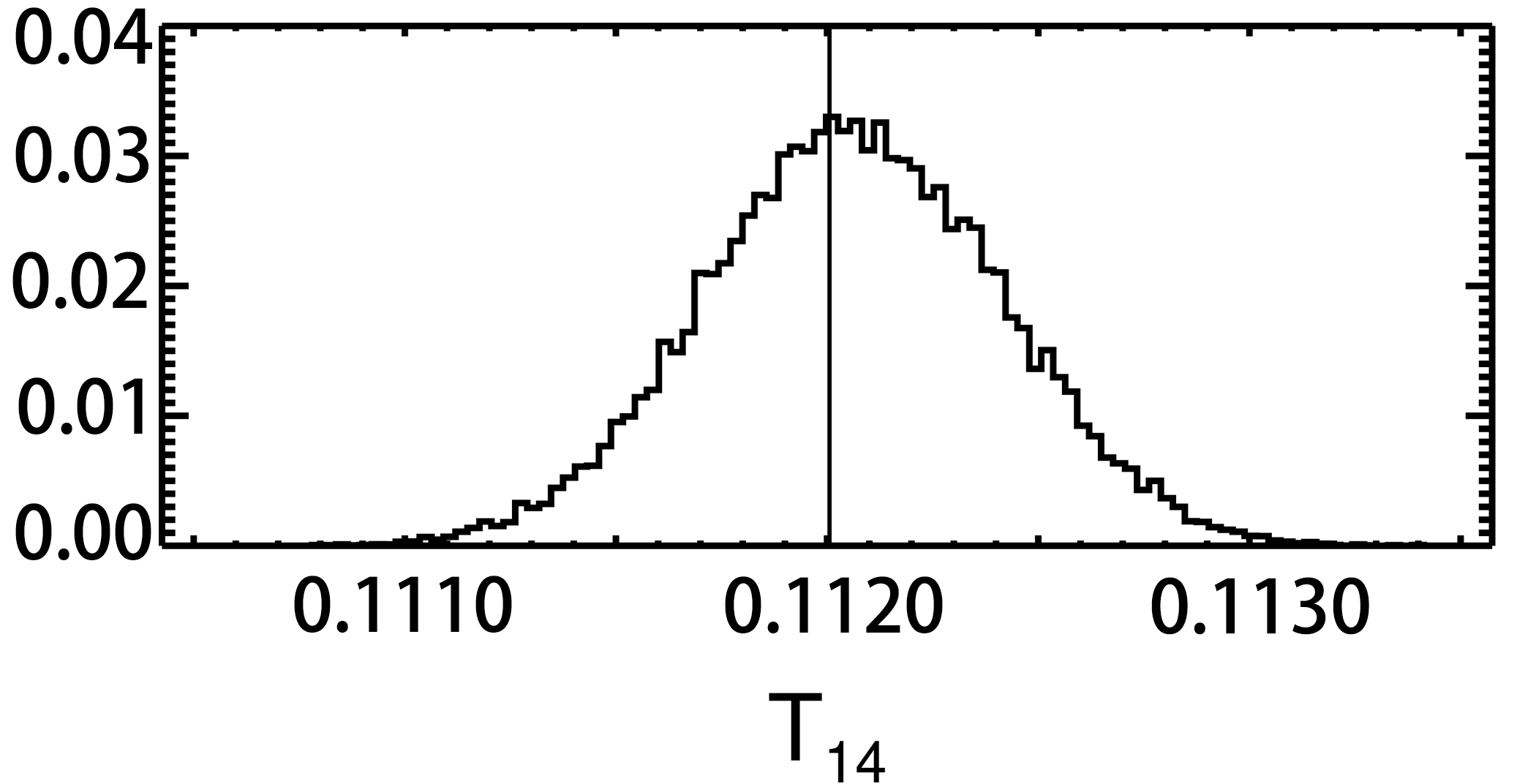}}
\subfigure[]{\includegraphics[width=55mm]{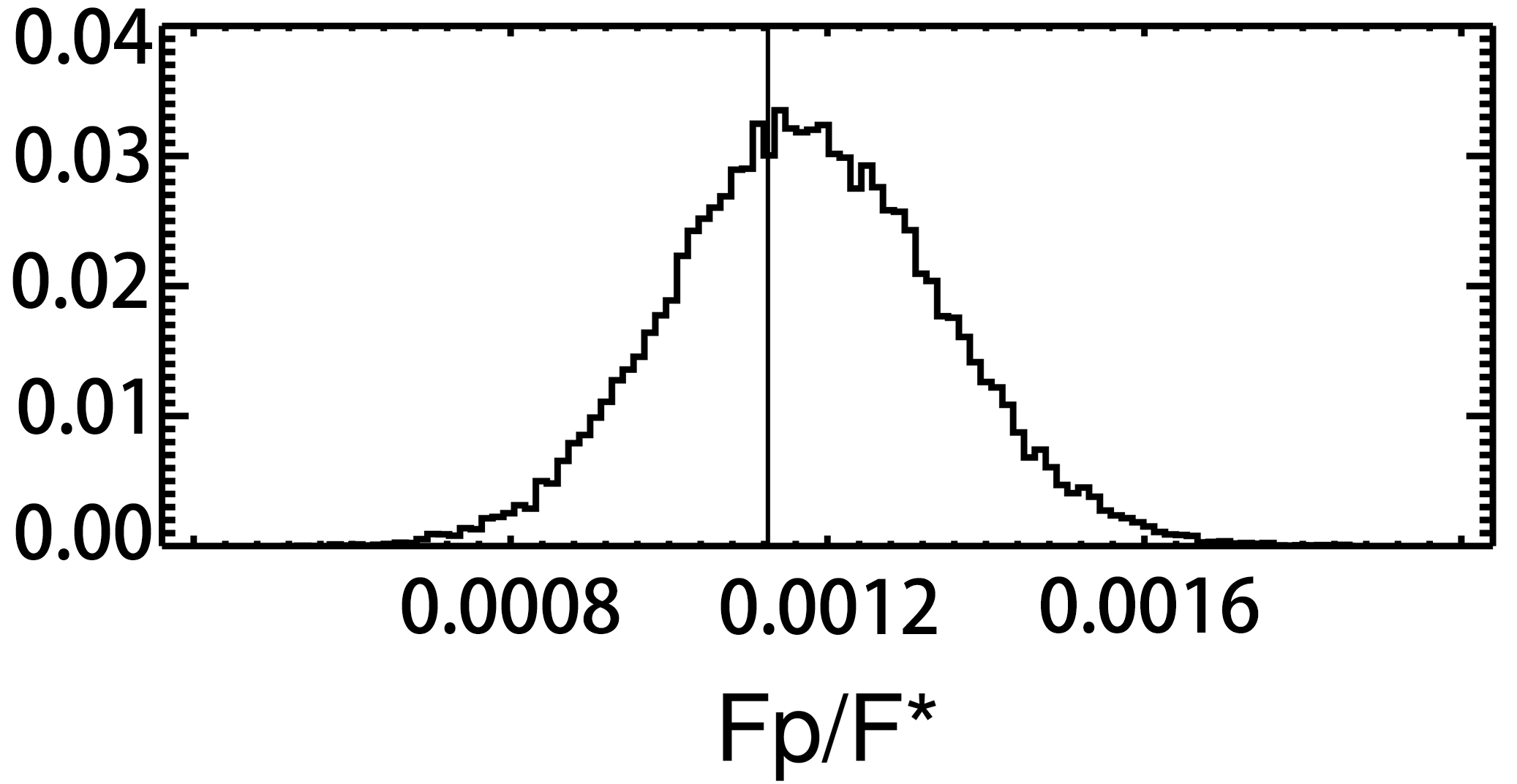}}

\caption{The distribution plots of the \texttt{EXOFAST} fitting parameters in $J$ band.The y-axis represents for probability.}

   \label{Fig:exofast}
\end{figure*}
    
\begin{figure*}
   \centering
  \includegraphics[width=16cm, angle=0]{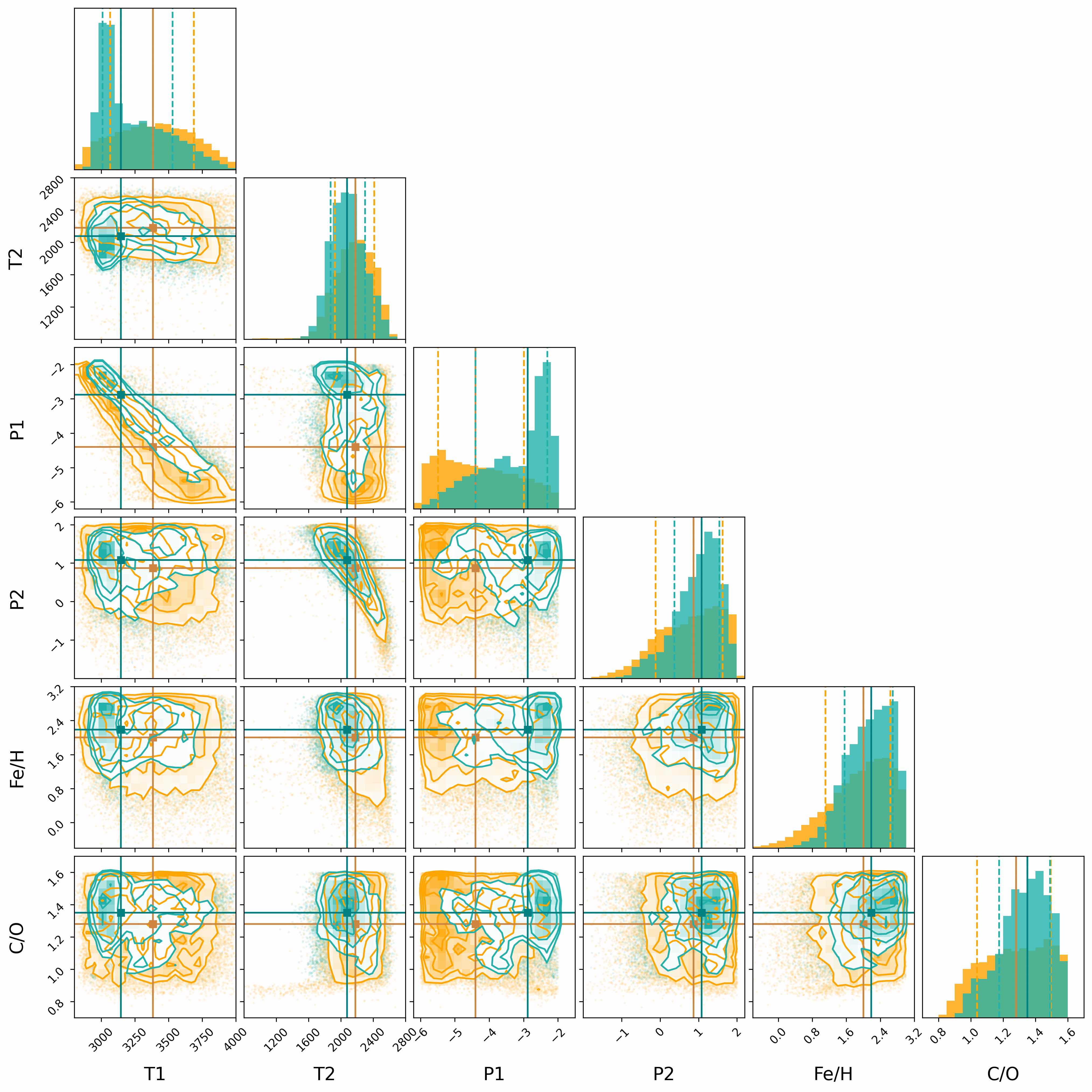}
   \caption{The retrieved atmospheric properties for the EQ retrievals on the unoffset data (green) and the ``os$=-200$'' data (yellow) with petitRADTRANS. The panels show the posterior distribution of parameters from the nested sampling run.  Darker shading corresponds to higher posterior probability. The diagonal shows a one-dimensional histogram for each parameter, with dotted lines denoting the median and 1$\sigma$ credible interval. The solid lines indicates the maximum likelihood estimation value.
   } 
   \label{Fig:eq_corner}
\end{figure*}

\begin{figure*}
   \centering
  \includegraphics[width=17cm, angle=0]{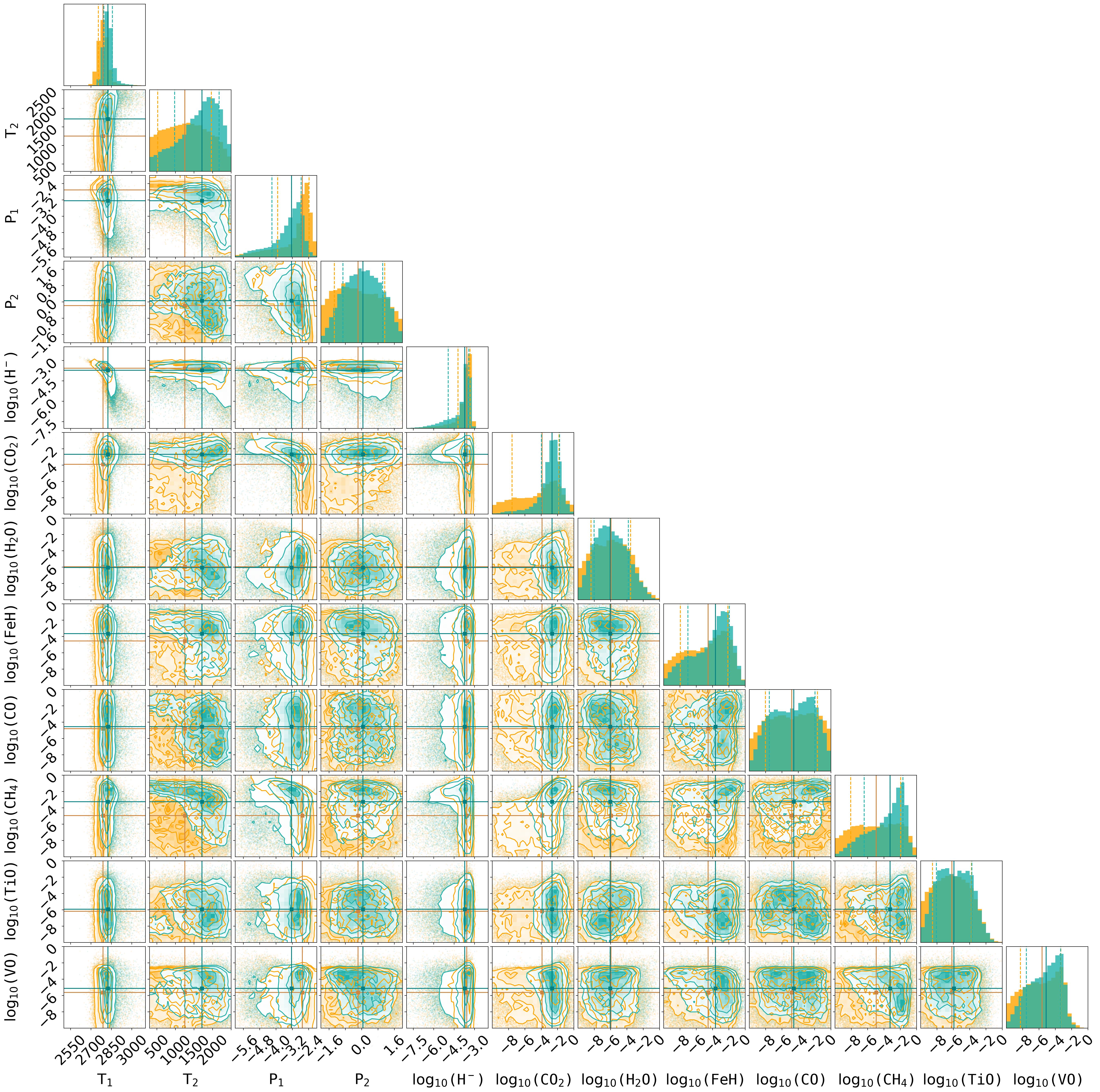}
   \caption{The retrieved atmospheric properties for the FREE retrieval on the unoffset data (green) and the ``os$=-50$'' data (yellow) with petitRADTRANS. The orange corners represent for the FREE retrieval after applying an offset of -50\,ppm to the HST data.} 
   \label{Fig:free_corner}
\end{figure*}


\bsp	
\label{lastpage}
\end{document}